\documentclass[useAMS,usenatbib]{mn2e}

\usepackage{amsmath}
\usepackage{amssymb}
\usepackage{graphicx}
\usepackage{graphics}
\usepackage{pdflscape}
\usepackage{float}
\usepackage{epstopdf}
\usepackage{multicol}
\usepackage{breakurl}
\usepackage{textcomp}
\usepackage{enumitem}
\usepackage{subfigure}
\usepackage[bf,labelsep=period,font=small]{caption}
\usepackage{subfloat}
\usepackage{subfigure}
\usepackage{gensymb}
\usepackage{supertabular}
\usepackage{longtable,ltxtable}
\usepackage{afterpage}
\usepackage[flushleft]{threeparttable}
\title[Follow-up of 8 candidate FK~Comae stars]{Spectropolarimetric follow-up of 8 rapidly rotating, X-ray bright FK~Comae candidates}
\author[J. Sikora et al.]
	{J.~Sikora,$^{1,}$\thanks{Based on observations obtained at the Canada-France-Hawaii Telescope (CFHT) which is operated by the National Research Council of Canada, the Institut National des Sciences de l'Univers of the Centre National de la Recherche Scientifique of France, and the University of Hawaii.} J.~Rowe,$^1$ S.~B.~Howell,$^2$ E.~Mason,$^3$ G.~A.~Wade$^4$\\
$^1$Department of Physics and Astronomy, Bishop's University, Sherbrooke, Qu{\'e}bec, Canada, J1M 1Z7\\
$^2$NASA Ames Research Center, Moffett Field, CA 94035, USA\\
$^3$INAF-OATS, Via G.B. Tiepolo, 11, I-34143, Trieste, Italy\\
$^4$Department of Physics and Space Science, Royal Military College of Canada, PO Box 17000 Kingston, Ontario, Canada, K7K 7B4\\}

\begin{document}

\date{Accepted 2020 May 12}

\pagerange{\pageref{firstpage}--\pageref{lastpage}} \pubyear{2020}

\maketitle

\label{firstpage}

\begin{abstract}
Our understanding of the evolved, rapidly rotating, magnetically active, and apparently single FK~Comae stars is significantly hindered by their extreme rarity: only two stars in addition to FK~Com itself are currently considered to be members of this class. Recently, a sample of more than 20 candidate FK~Comae type stars was identified within the context of the \emph{Kepler-Swift} Active Galaxies and Stars (KSwAGS) survey. We present an analysis of high-resolution Stokes $V$ observations obtained using ESPaDOnS@CFHT for 8 of these candidates. We found that none of these targets can be considered members of the FK~Comae class based primarily on their inferred rotational velocities and on the detection of spectroscopic binary companions. However, 2 targets show evidence of magnetic activity and have anomalously high projected rotational velocities ($v\sin{i}$) relative to typical values associated with stars of similar evolutionary states. EPIC~210426551 has a $v\sin{i}=209\,{\rm km\,s}^{-1}$, an estimated mass of $1.07\,M_\odot$, and, based in part on its derived metallicity of ${\rm [M/H]}=-0.4$, it is either an evolved main sequence (MS) star or a pre-MS star. KIC~7732964 has a mass of $0.84\,M_\odot$, lies near the base of the red giant branch, and exhibits a $v\sin{i}=23\,{\rm km\,s}^{-1}$. We find that these two objects have similar characteristics to FK~Com (albeit less extreme) and that their rapid rotation may be inconsistent with that predicted for a single star evolutionary history. Additional observations are necessary in order to better constrain their evolutionary states and whether they have short-period binary companions.
\end{abstract}


\section{Introduction}\label{sect:intro}

FK~Comae type stars are amongst the most unusual non-degenerate stars that are currently known. These apparently single, red giant branch stars are defined largely by their high rotational velocities $\gtrsim100\,{\rm km\,s}^{-1}$ \citep{bopp1981}. FK~Com itself has a projected rotational velocity of $v\sin{i}\approx160\,{\rm km\,s}^{-1}$ \citep{huenemoerder1993}, which is particularly surprising considering that the majority of evolved stars have $v\sin{i}<10\,{\rm km\,s}^{-1}$ \citep[e.g.][]{demedeiros1996,cutispoto2002}. FK~Com also exhibits enhanced magnetic activity as evidenced by the detection of (1) strong and variable H$\alpha$ and Ca~{\sc ii} H and K emission \citep{herbig1958,ramsey1981}, (2) highly energetic flares \citep{jetsu1991,oliveira1999}, (3) a mean longitudinal magnetic field ranging from $\langle B_z\rangle\approx60-272\,{\rm G}$ \citep{korhonen2009}, and (4) strong UV and X-ray emission \citep{bopp1981,ayres2016}.

The FK~Comae type stars are extremely rare since only two other members (ET~Dra and HD~199178) have been discovered. A number of apparently single, evolved stars have been found with lower yet still anomalously high $v\sin{i}$ values of $10-80\,{\rm km\,s}^{-1}$ \citep[e.g.][]{demedeiros1996,drake2002,carlberg2011}. As with FK~Com, these rapid rotators appear to challenge our understanding of how the rotation rates of single stars evolve over time \citep[e.g.][]{bouvier1997,spada2011}. One theory that has been proposed to explain these unexpectedly high rotation rates involves the transport of internal angular momentum from the star's rapidly rotating core during the first dredge-up phase \citep{simon1989,fekel1993}; however, \citet{privitera2016b} argue that the acceleration of the star's surface provided by such a scenario is largely insignificant. It has also been proposed that, as first outlined by \citet{webbink1976a}, the observed rapid rotation rates may be a byproduct of stellar merger events \citep[e.g.][]{bopp1981,strassmeier1998,dasilva2015}. In order to better understand how the rapid rotation rates of FK~Comae type stars are produced, the small sample size likely needs to be increased.

\begin{table*}
	\caption{Spectral types, details of the obtained ESPaDOnS observations, derived radial velocities, and derived longitudinal magnetic field strengths. Columns 1 to 4 list each target's identifier, spectral type, $V$ magnitude, and the maximum S/N per pixel near $4\,000\,{\rm \AA}$ and $8\,000\,{\rm \AA}$. Columns 5 to 7 list the HJDs, the $v_r$ values of the primary and secondary (if a secondary component was detected) components. Columns 5 and 6 list the derived $\langle B_z\rangle$ values and the detection status (DD and ND corresponding to definite detection and no detection, respectively).}
	\label{tbl:obs_vr_Bz}
	\begin{center}
	\begin{tabular}{@{\extracolsep{\fill}}l c c c c c c c r@{\extracolsep{\fill}}}
		\hline
		\hline
		\noalign{\vskip0.5mm}
				ID  & Sp.  & $V$   & S/N          & Obs. HJD       & $v_{r,{\rm A}}$ & $v_{r,{\rm B}}$ & $\langle B_z\rangle$ & Detect. \\
				    & Type & (mag) & (pxl$^{-1}$) & $-2\,458\,000$ & (km$\,s^{-1}$)  & (km$\,s^{-1}$)  & (G)                  & Status  \\
				(1) & (2)  & (3)   & (4)          & (5)            & (6)             & (7)             & (8)                  & (9)     \\
		\noalign{\vskip0.5mm}
		\hline
		\noalign{\vskip0.5mm}
EPIC~210355746 &                                       &  9.8 &  50/290 & 476.754 &   $8.3\pm0.5$ &               &  $-1.2\pm6.2$ & ND~ \\
               &                                       &      &  50/300 & 564.719 &   $8.5\pm0.4$ &               &  $-2.8\pm6.3$ & ND~ \\
EPIC~210384590 & G0$^{\rm a\,}$                        &  9.0 &  70/400 & 476.778 &  $23.7\pm0.1$ &               &   $5.8\pm1.9$ & DD~ \\
EPIC~210426551 &                                       & 10.3 &  40/280 & 477.779 &      $10\pm4$ &               &   $493\pm335$ & ND~ \\
EPIC~210505125 & G0$^{\rm a\,}$                        &  8.2 &  90/420 & 413.118 & $-40.2\pm0.3$ &  $16.6\pm0.5$ &               & ND~ \\
               &                                       &      &  80/380 & 441.887 &   $7.8\pm0.3$ & $-33.1\pm0.3$ &               & ND~ \\
EPIC~220533366 & K0$^{\rm a\,}$                        &  8.8 &  50/470 & 413.826 & $-49.5\pm0.7$ &               & $-24.2\pm3.5$ & DD~ \\
               &                                       &      &  30/410 & 475.772 & $-20.4\pm0.6$ &               & $-21.9\pm3.5$ & DD~ \\
KIC~4857678    & F3V$^{\rm b}$, F2-5III$^{\rm c\,}$    &  7.0 & 120/480 & 413.782 &     $-22\pm1$ &               &    $-58\pm68$ & ND~ \\
               &                                       &      &  20/100 & 441.738 &     $-22\pm2$ &               &   $243\pm327$ & ND~ \\
KIC~6150124    & K0$^{\rm a}$, G0-9III$^{\rm c\,}$     &  7.4 &  80/590 & 413.793 &  $10.6\pm0.2$ &               &   $1.0\pm1.0$ & ND~ \\
KIC~7732964    & G6-7IV-III$^{\rm c\,}$                & 11.1 &  30/210 & 645.828 & $-18.4\pm0.9$ &               & $-9.4\pm10.0$ & DD$^\dagger$ \\
		\noalign{\vskip0.5mm}
		\hline \\
\noalign{\vskip-0.3cm}
\multicolumn{8}{l}{$^{\rm a\,}$\citet{cannon1993}, $^{\rm b\,}$\citet{frasca2016}, $^{\rm c\,}$\citet{howell2016}, $^\dagger$Signal also detected in diagnostic null.} \\
	\end{tabular}
	\end{center}
\end{table*}

As part of the \emph{Kepler-Swift} Active Galaxies and Stars (KSwAGS) survey \citep{smith2015}, simultaneous X-ray and UV observations were obtained for four modules of the \emph{Kepler} Field of View (FOV) ($\sim6$ square degrees). These modules were surveyed in a series of tiled $2~{\rm ks}$ observations using the \emph{Swift} X-ray Telescope (XRT), which obtains both X-ray and UV images with an angular resolution of 18 arcseconds \citep{gehrels2004}. Ninety-three X-ray sources were detected at signal-to-noise ratios (S/N) $>3$ (i.e. 12 counts or $0.006~{\rm cts\,s^{-1}}$ at the mean exposure time and background level of the snapshots).

The detected sources have X-ray to optical flux ratios exceeding 100 times the maximum value observed for the Sun. UV counterparts were also detected for 60 of the 93 X-ray sources. On the presumption that many of these sources may be FK~Com candidates, \citet{howell2016} carried out a study of 20 of the 60 X-ray and UV-bright sources with the primary goal of constraining these stars' rotational velocities. Based on the available \emph{Kepler} light curves and newly obtained medium-resolution optical spectra ($R\sim7\,700-10\,000$), \citet{howell2016} concluded that 18 of the 20 stars in their sample are apparently single, evolved, and rapidly rotating; their results are therefore consistent with the notion that a significant fraction of the 60 sources identified by the KSwAGS survey are similar in nature to FK~Com and thus, they may be consistent with the predicted properties of stellar merger products.

The KSwAGS survey has continued to identify X-ray sources within several fields of the \emph{K2} mission \citep{howell2014} based on newly obtained and archival measurements (K. Smith, private communication). These ongoing efforts have involved \emph{Swift} XRT measurements, obtained using a similar observing strategy to that described by \citet{smith2015}, along with archival Chandra measurements and measurements obtained by the XMM-Slew Survey and the ROSAT All-Sky Survey. Low-resolution ($R\sim500-4\,000$) optical follow-up spectra were obtained for those sources having $V\lesssim17\,{\rm mag}$ using the DeVeny spectrograph installed at the Discovery Channel Telescope at Lowell Observatory. From these spectra, no line measurements or luminosity class estimates were attempted; however, using standard MK spectral classification techniques \citep[see e.g.,][]{keenan1941,garrison1984}, which make use of the shape and extent of the Balmer jump, the width and depth of broad molecular bands, and the general shape of the relative flux distribution, approximate spectral types were estimated. These rough classifications were not used to develop any of the stellar parameters used herein, but gave credence to the fact that the stars are FGK types and eyeball proof that the temperatures and derived luminosity classes reported in the \emph{Gaia} DR2 catalog \citep{andrae2018} are sensible.

In this study, we present new high-resolution Stokes $V$ observations obtained using ESPaDOnS installed at the Canada France Hawaii Telescope (CFHT) for 8 stars that have been identified as FK~Comae type candidates within the broader context of the KSwAGS survey. Three of these targets are discussed by \citet{howell2016} while the remaining 5 targets were identified as bright X-ray sources in Fields 4 and 8 of the \emph{K2} mission. The primary goals of our study are to (1) precisely measure each target's $v\sin{i}$ value in order to identify or confirm their (presumed) rapid rotation rates, (2) search for evidence of binary companions (via radial velocity variations) that may provide an alternative explanation for the primary component's rapid rotation (i.e. that the primary component has been spun-up by tidal interactions rather than by a stellar merger event), and (3) search for Zeeman signatures that are indicative of magnetic activity and organized magnetic fields. In Section \ref{sect:obs}, we discuss the new spectroscopic observations that were obtained along with the publicly available \emph{Kepler} and \emph{K2} light curves that are available for each target. The magnetic field and radial velocity measurements that we derive are presented in Section \ref{sect:lsd}. In Section \ref{sect:spec}, we present our spectral modelling analysis from which $v\sin{i}$ and other fundamental parameters are derived. In Section \ref{sect:hrd}, each target is compared with evolutionary model grids. Our analysis of the \emph{Kepler} and \emph{K2} light curves is presented in Section \ref{sect:LC} in which we search for plausible signatures of rotational modulation. Finally, in Sections \ref{sect:disc} and \ref{sect:summ} we provide a discussion and a summary of our results.

\section{Observations}\label{sect:obs}

The study presented here makes use of newly obtained spectropolarimetric measurements along with publicly available \emph{Kepler} and \emph{K2} light curves. Here we discuss those data sets.

\subsection{ESPaDOnS Spectra}\label{sect:esp_obs}

The 8 targets included in this study were observed using the high-resolution ($R\sim65\,000$) \'echelle spectropolarimeter, ESPaDOnS, installed at CFHT. This instrument provides polarized (circular and linear) and unpolarized spectra spanning a wavelength range of approximately $3\,600-10\,000\,{\rm \AA}$.

Circularly polarized (Stokes $V$) spectra were acquired using ESPaDOnS between Oct.~21,~2018 and June~11,~2019. The observing strategy involved obtaining 2 measurements separated in time by more than 1 week for each target. Four of the targets (EPIC~210355746, EPIC~210505125, EPIC~220533366, and KIC~4857678) were observed twice during this time span; these observations were separated in time by approximately 1 to 3 months. Due to scheduling and/or weather constraints, the remaining targets (EPIC~210384590, EPIC~210426551, KIC~6150124, and KIC~7732964) were only observed once. The observations have S/Ns per CCD pixel ranging from as $\approx20-120$ near a wavelength of $\lambda=4\,000{\rm \AA}$ to $\approx100-590$ near $\lambda=8\,000\,{\rm \AA}$.

All of the observations were reduced using the Upena pipeline feeding the {\sc libre-esprit} reduction software described by \citet{donati1997}. The pipeline automatically applies wavelength corrections to account for the earth's rotation and orbital motion and is capable of automatically normalizing the observations. We opted to manually normalize the spectra by fitting a polynomial (typically of order 1 or 2) to the continuum of each (unnormalized) spectral order. In Table \ref{tbl:obs_vr_Bz}, we list each target's $V$ magnitude, spectral type reported in the literature (if available), along with the exposure times, dates, and maximum S/Ns of the obtained observations.

\subsection{\emph{Kepler} and \emph{K2} Light Curves}\label{sect:LC_obs}

Three of the 8 targets in our sample are located in the original \emph{Kepler} FOV and 5 are located in Fields 4 and 8 of the \emph{K2} mission. Optical light curves obtained with the \emph{Kepler} spacecraft are publicly available for all of our targets. These light curves were downloaded from the Mikulski Archive for Space Telescopes (MAST)\footnote{https://archive.stsci.edu}. All of the light curves have a cadence of $30\,{\rm min}$. The \emph{K2} light curves span a time period of $\approx2.5\,{\rm months}$ (EPIC~210355746, EPIC~210384590, EPIC~210426551, EPIC~210505125, and EPIC~220533366) while the 3 \emph{Kepler} light curves span either $2.4\,{\rm yrs}$ (KIC~4857678) or $4\,{\rm yrs}$ (KIC~6150124 and KIC~7732964).

We used the ``presearch data condition simple aperture photometry" (PDC\_SAP) light curves, which have been processed in such a way as to remove systematic errors including outliers, systematic trends, and instrumental signals \citep{smith2012}. We carried out additional post-processing consisting of outlier removal and basic detrending in which a first order polynomial was fit to the full light curve and the flux measurements divided by the resulting fit.

\section{Least-Squares Deconvolution profiles}\label{sect:lsd}

\begin{figure*}
	\centering
	\begin{minipage}{0.8\columnwidth}
		\centering
		\subfigure{\includegraphics[width=1.0\columnwidth]{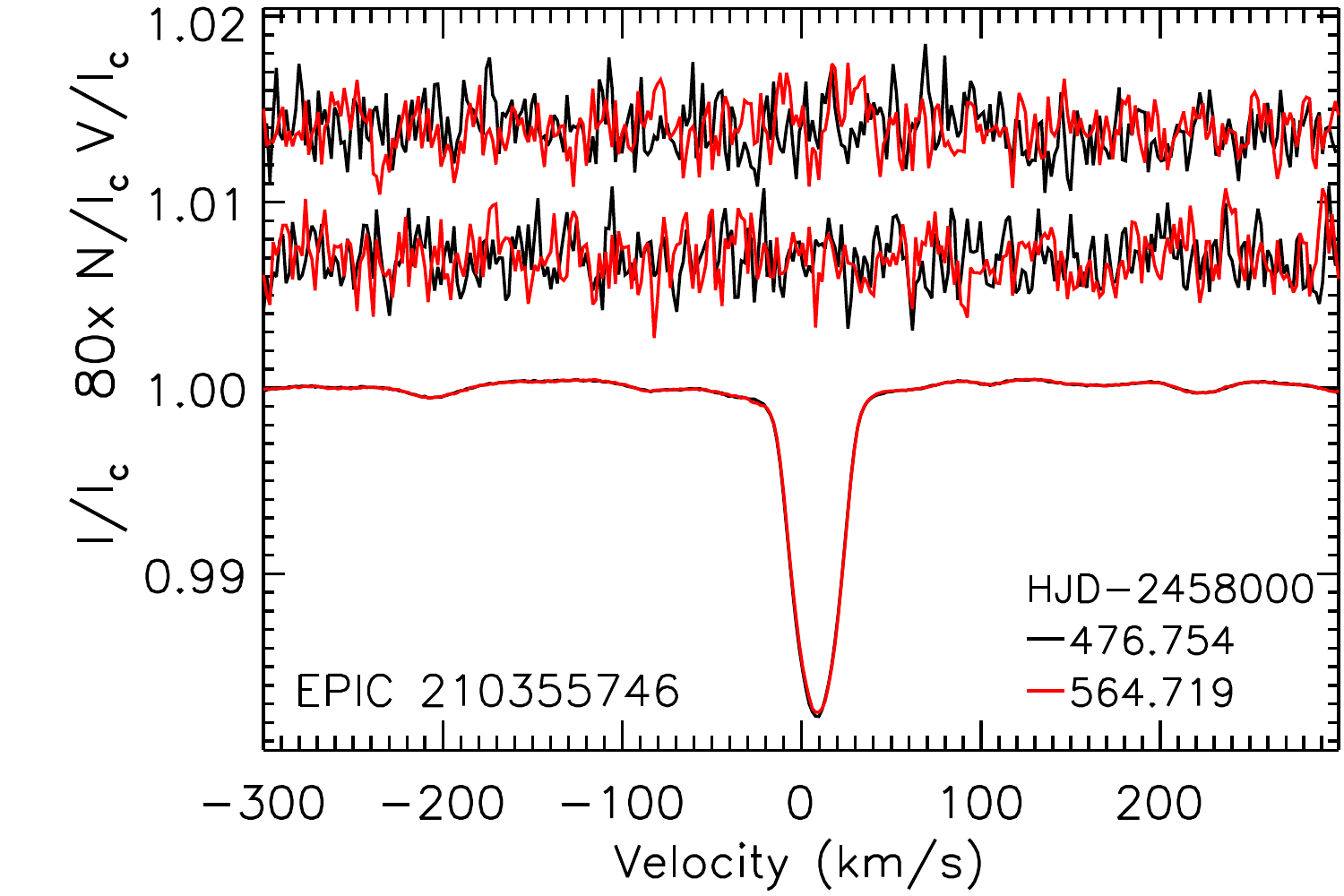}}\vspace{-0.4cm}
		\subfigure{\includegraphics[width=1.0\columnwidth]{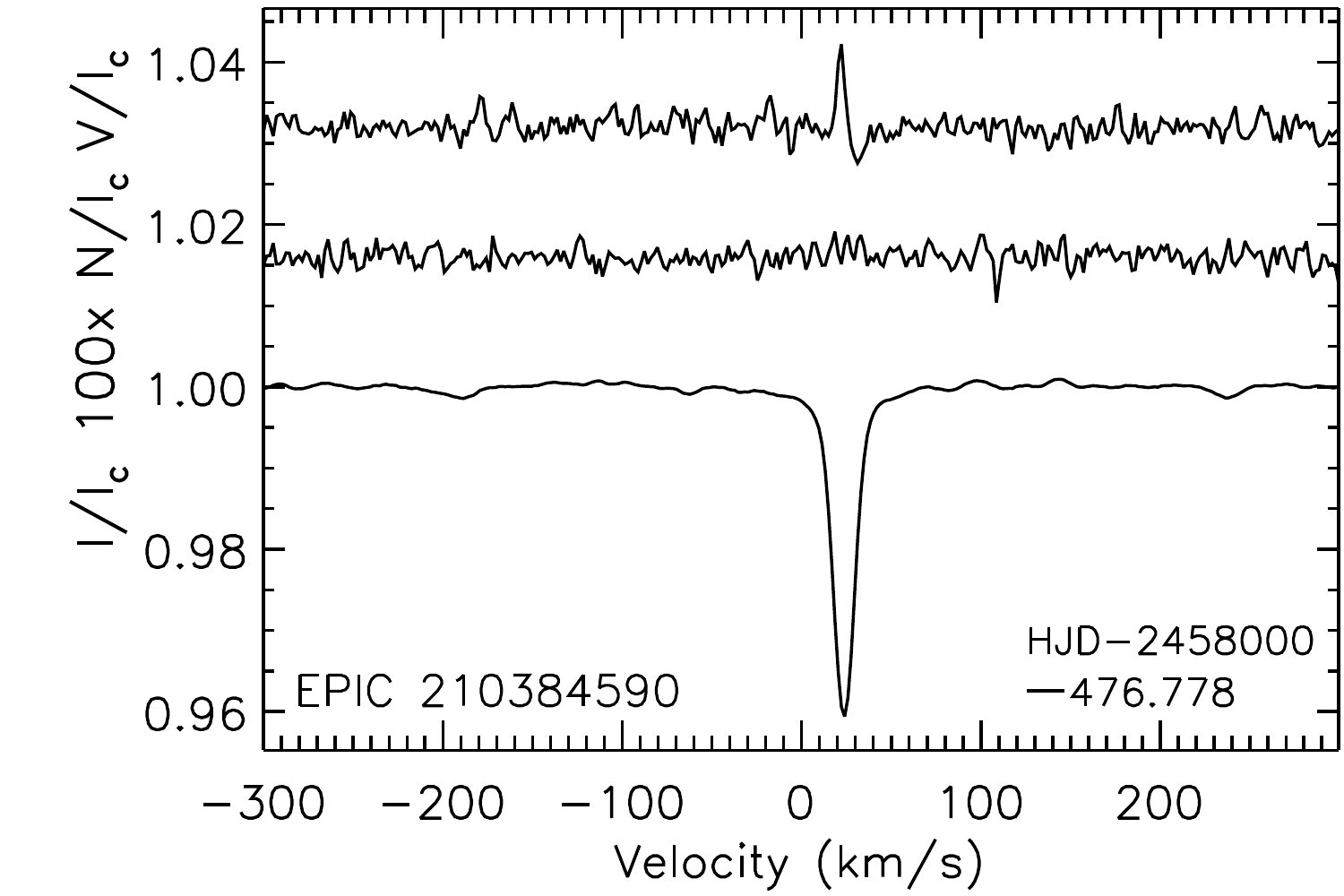}}\vspace{-0.4cm}
		\subfigure{\includegraphics[width=1.0\columnwidth]{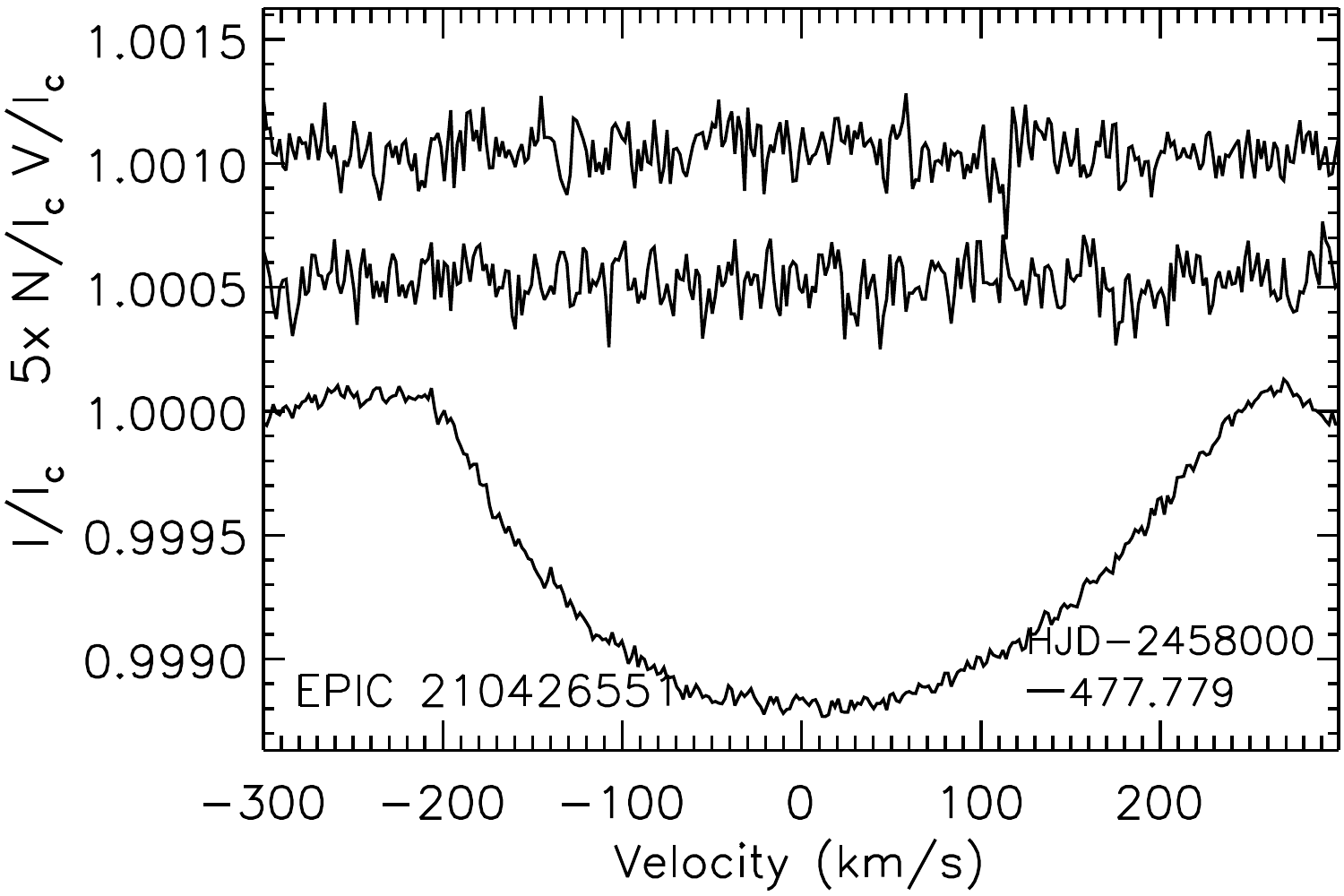}}\vspace{-0.4cm}
		\subfigure{\includegraphics[width=1.0\columnwidth]{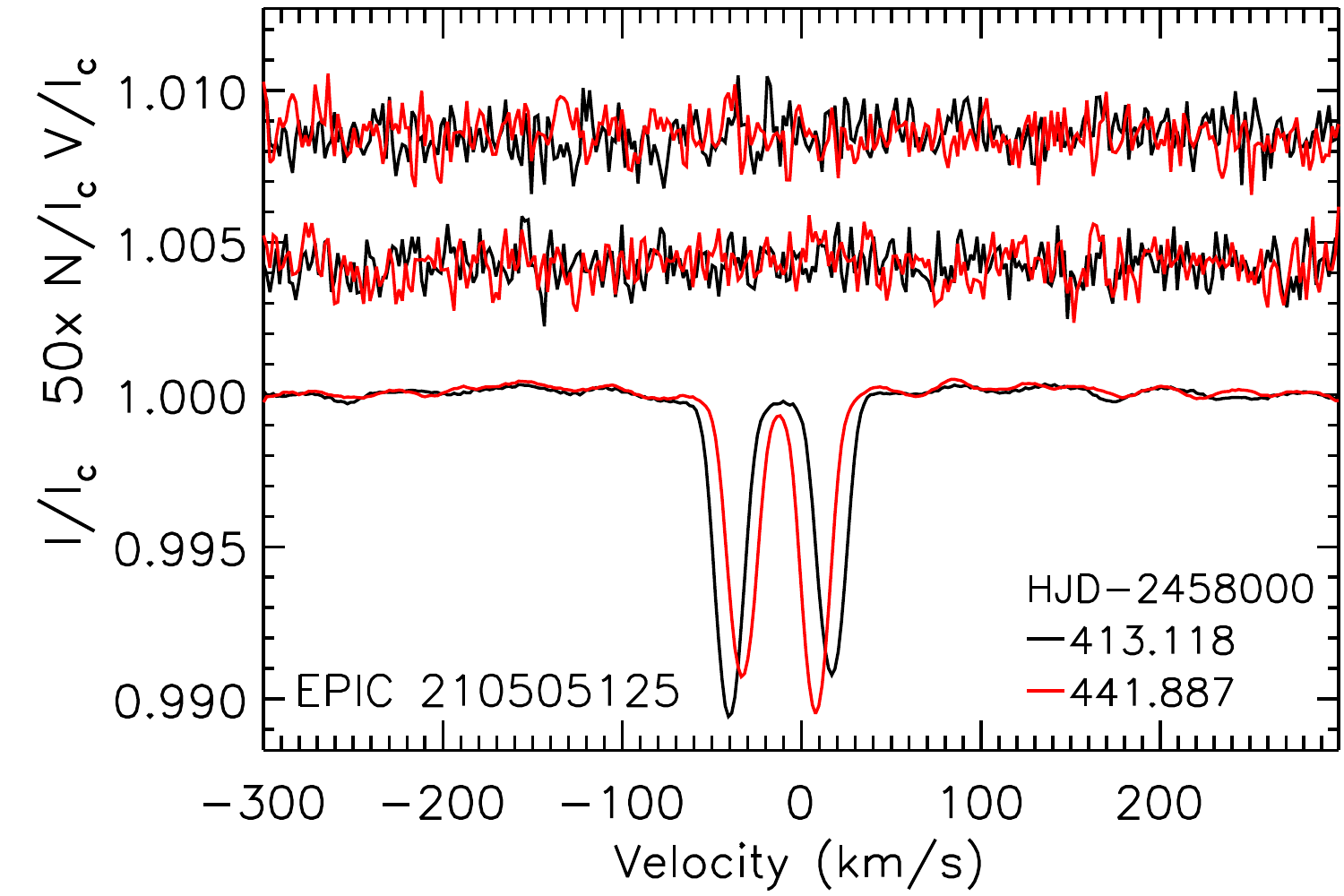}}
	\end{minipage}
	\begin{minipage}{0.8\columnwidth}
		\centering
		\subfigure{\includegraphics[width=1.0\columnwidth]{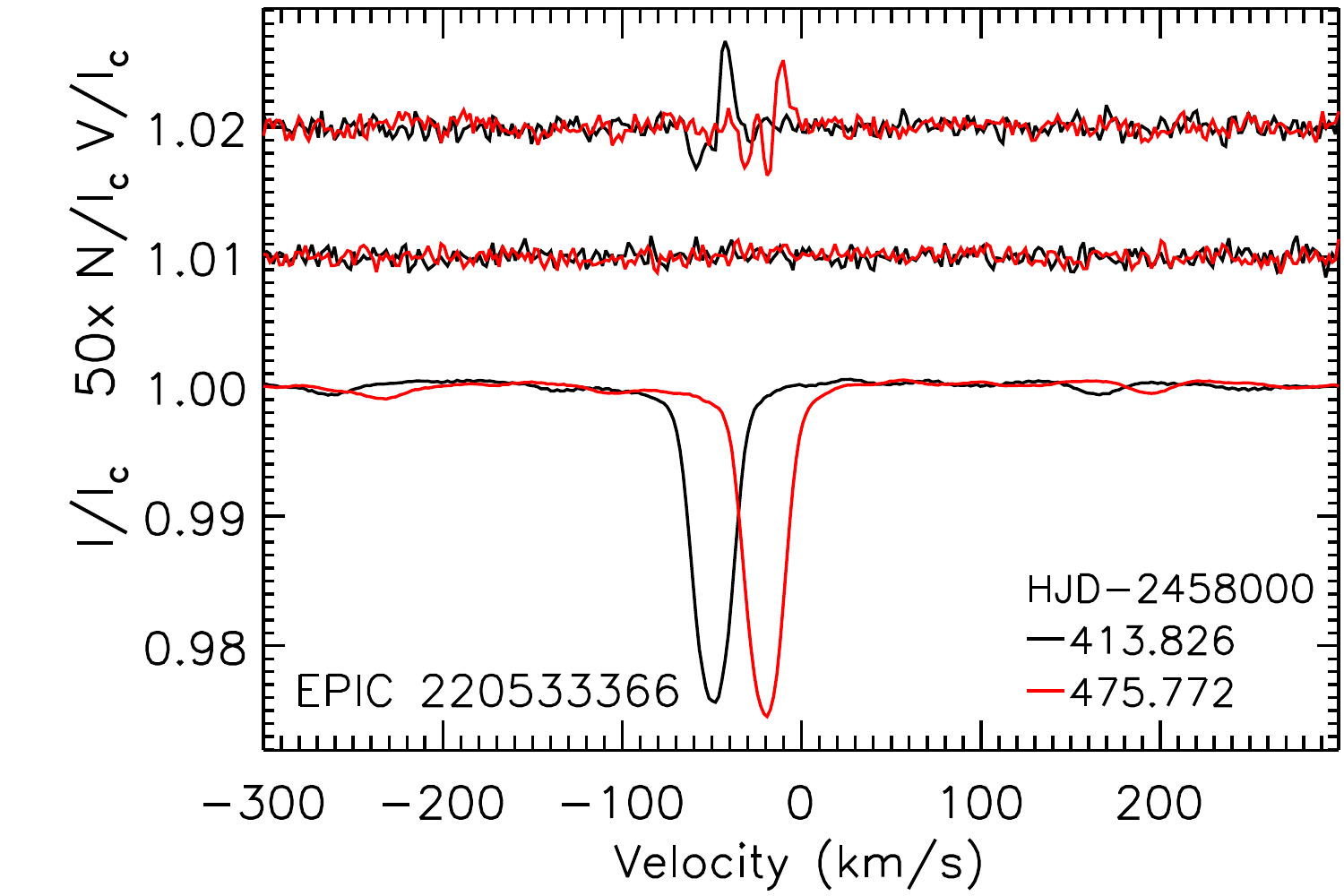}}\vspace{-0.4cm}
		\subfigure{\includegraphics[width=1.0\columnwidth]{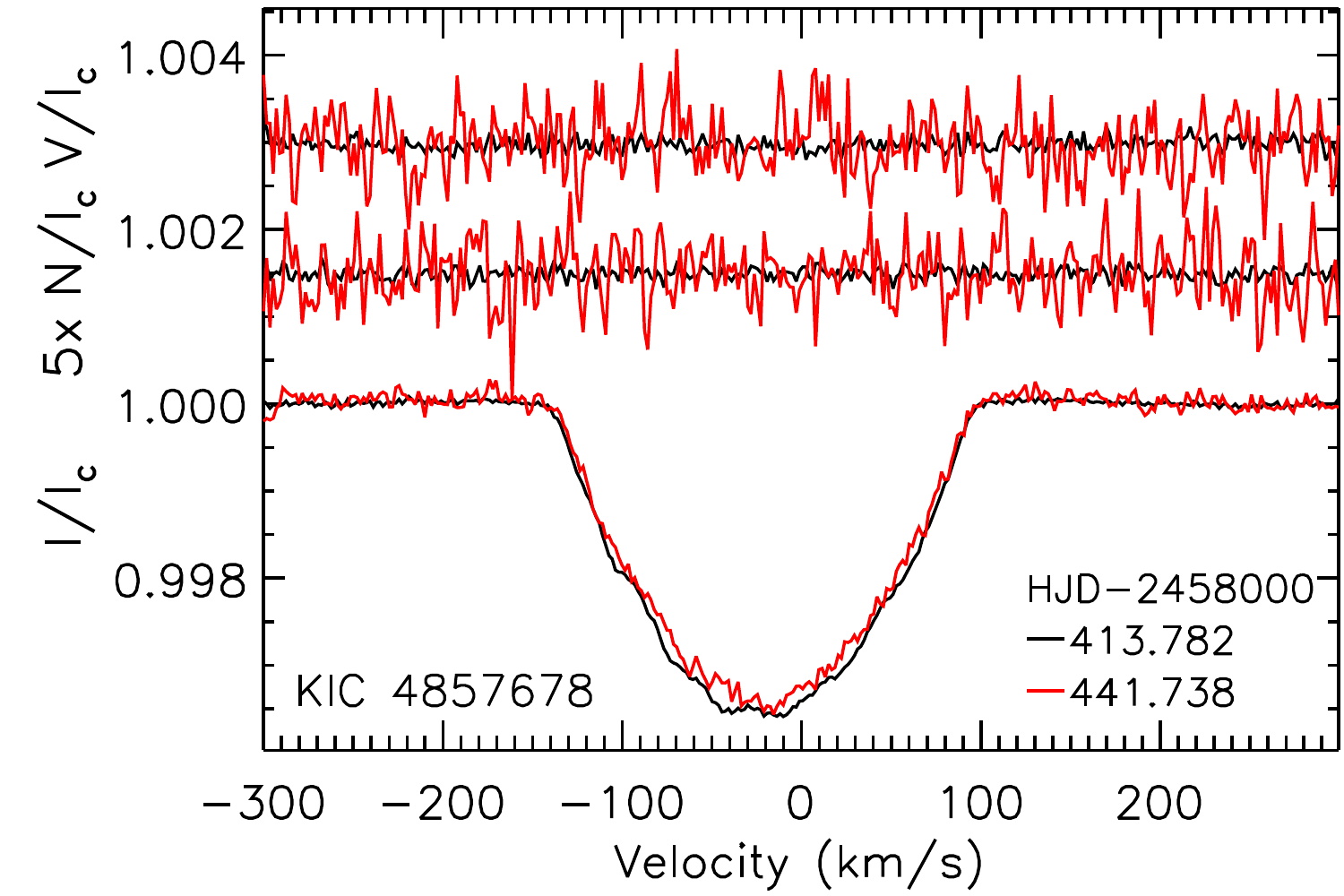}}\vspace{-0.4cm}
		\subfigure{\includegraphics[width=1.0\columnwidth]{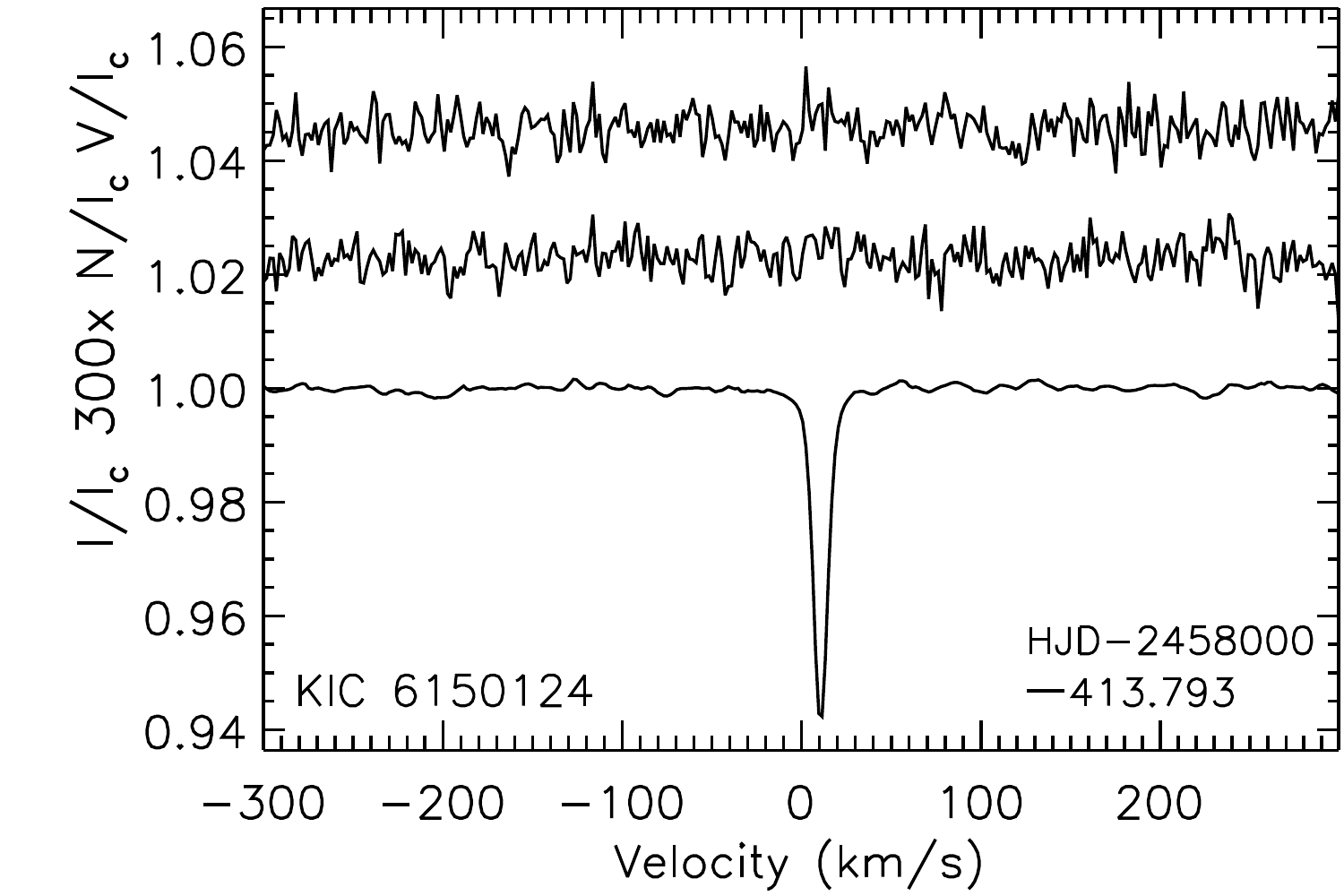}}\vspace{-0.4cm}
		\subfigure{\includegraphics[width=1.0\columnwidth]{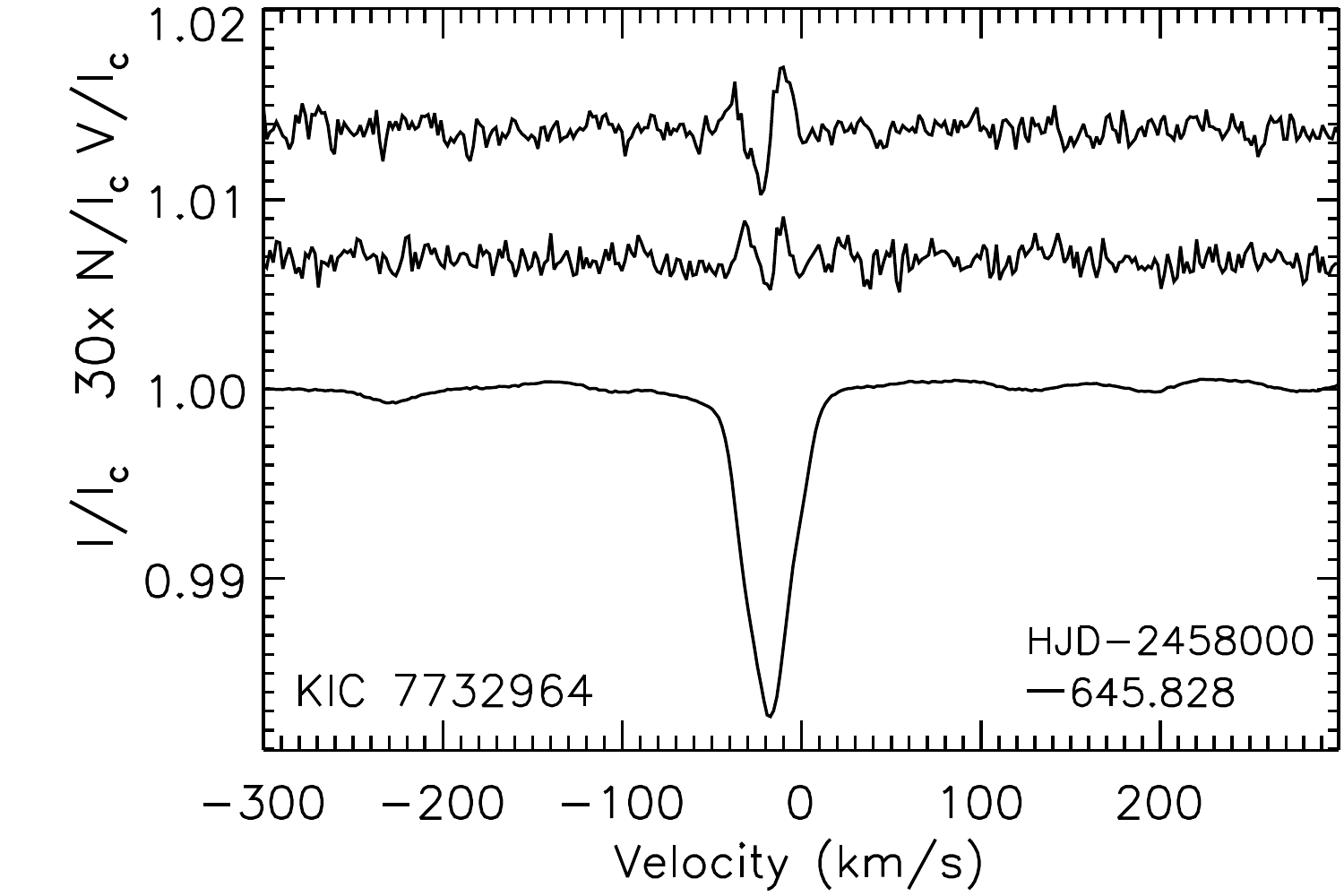}}
	\end{minipage}
	\caption{LSD pseudo-line profiles derived from the ESPaDOnS observations of the 8 targets included in this study. In each panel, the normalized Stokes $V$ and $I$ profiles are plotted ($V/I_c$ and $I/I_c$; top and bottom, respectively) along with the diagnostic null ($N/I_c$; middle). Note that $V/I_c$ and $I/I_c$ have been shifted vertically and scaled for visual clarity. The appearance of positive and negative lobes in $V/I_c$ correspond to circularly polarized Zeeman signatures produced by magnetic fields that are visible at the star's surface. Similar features appearing in the diagnostic null (e.g. KIC~7732964) indicate that the detected Zeeman signatures may be instrumental in origin. Spectroscopic binary companions are clearly detected in $I/I_c$ based on radial velocity variations (EPIC~220533366) and/or the presence of multiple components (EPIC~210505125).}
	\label{fig:all_lsd}
\end{figure*}

We generated Least-Squares Deconvolution (LSD) profiles \citep{donati1997,kochukhov2010a} for each of the measurements obtained using ESPaDOnS. LSD is a cross-correlation technique that combines a large number of similarly-shaped spectral lines in order to produce a pseudo line profile with a significanty higher S/N. We used the resulting Stokes $I$ and $V$ LSD profiles to (i) derive the radial velocities of all stellar components visible in the spectra and (ii) identify circularly polarized Zeeman signatures and derive the associated mean longitudinal field strengths ($\langle B_z\rangle$).

The LSD technique requires a line mask consisting of the wavelengths and relative depths of spectral lines that are expected to be found in the observed spectra based primarily on the star's effective temperature ($T_{\rm eff}$) and surface gravity ($\log{g}$). These masks were constructed using line lists generated by the `Extract Stellar' tool provided by the Vienna Atomic Line Database (VALD) \citep{ryabchikova2015}. The specified $T_{\rm eff}$ values were taken from the \emph{Gaia} DR2 catalog \citep{andrae2018} and we adopted $\log{g}$ values of $3.0$ or $4.0\,{\rm (cgs)}$ based on each star's position in the Hertzsprung-Russell Diagram (presented in Sect. \ref{sect:hrd}). All lines with wavelengths within the broad wings of the Balmer lines (i.e. within approximately $30\,{\rm \AA}$ of the central Balmer line wavelength) were removed from the line masks along with the Balmer lines themselves and those lines that were found to be contaminated with tellurics. The Stokes $I$, Stokes $V$, and the two diagnostic null\footnote{The diagnostic null fluxes obtained during each measurement are used to identify polarized signals that may be instrumental in origin \citep[e.g.][]{donati1997}.} ($N_1$ and $N_2$) LSD profiles were then computed using the normalized spectra and the generated line masks. In Fig. \ref{fig:all_lsd}, we show all of the Stokes $I$ and $V$ LSD profiles associated with the stars in our sample.

\subsection{Radial velocities and binarity}\label{sect:rv_bin}

Based on the computed LSD profiles, only one target (EPIC~210505125) was found to consist of more than one spectral component. EPIC~210505125 exhibits two components that are similar both in terms of the widths and depths of their Stokes $I$ profiles. Clear radial velocity shifts are apparent between the two observations of EPIC~220533366 suggesting that it consists of at least one additional dimmer (relative to the primary component) companion.

We derived the radial velocities ($v_r$) associated with all of the stellar components that were detected by fitting Gaussian functions to the Stokes $I$ profiles. Uncertainties in the derived $v_r$ values were estimated by bootstrapping the residuals in order to generate 500 bootstrapped data sets (see Sikora et al., submitted). We note that, in general, the Gaussian fits to the Stokes $I$ profiles are of reasonably high quality. The largest residuals are associated with KIC~7732964, which exhibits moderate asymmetries; the four Stokes $I$ sub-exposures used to obtain the Stokes $V$ measurement reveal variability that affects the entire profile over a time period of $37\,{\rm min}$, which suggests that the asymmetries in the averaged Stokes $I$ profile is produced by pulsations. The resulting $v_r$ values and their $3\sigma$ uncertainties derived for all of the targets in our sample are listed in Table \ref{tbl:obs_vr_Bz}.

\subsection{Longitudinal magnetic field}\label{sect:Bz}

In Fig. \ref{fig:all_lsd}, clear circularly polarized Zeeman signatures, which are manifest as positive and negative lobes (i.e. systematic deviations from zero) in the Stokes $V$ pseudo-line profiles, are detected for 3 of the 8 targets (EPIC~210384590, EPIC~220533366, and KIC~7732964). In the case of KIC~7732964, however, the same characteristic signal is visible in one of the two diagnostic null LSD profiles. This suggests that the detected Zeeman signature may not be instrinsic to the star but rather an observational artifact. Additional observations are required to confirm the presence of a detectable magnetic field at the star's surface.

We measured $\langle B_z\rangle$ from each of the observations using the computed Stokes $I$ and $V$ profiles using Eqn. 1 of \citet{wade2000}. We used mean wavelengths ($\approx480\,{\rm nm}$) and mean Land\'e factors ($\approx1.2$) that were calculated from the line masks discussed above in Sect. \ref{sect:lsd}. Integration ranges were adopted such that they span the width of the Stokes $I$ profiles (e.g. for EPIC~220533366, we used a width of $45\,{\rm km\,s}^{-1}$; for EPIC~210426551, we used a width of $383\,{\rm km\,s}^{-1}$). In Table \ref{tbl:obs_vr_Bz}, we list the derived $\langle B_{\rm z}\rangle$ measurements, their associated uncertainties, and the Zeeman signature detection probability \citep[no detection or definite detection as specified by][]{donati1997}.

\section{Spectral Modelling}\label{sect:spec}

The high-resolution spectra obtained for this study can be used to constrain various fundamental parameters associated with the sample stars including $T_{\rm eff}$, $\log{g}$, metallicity ([M/H]), the projected rotational velocity ($v\sin{i}$), the microturbulence ($\xi$), along with individual chemical abundances. We carried out a spectroscopic modelling analysis based largely on that described by Sikora et al. (submitted). This involved using the Grid Search in Stellar Parameters ({\sc gssp}) code published by \citet{tkachenko2015}. The {\sc gssp} code generates grids of synthetic spectra computed from {\sc LLmodels} atmospheric models \citep{shulyak2004}. The grids can be defined such that they span various ranges of the input parameters with associated grid resolutions.

\begin{figure}
	\centering
	\subfigure{\includegraphics[width=1\columnwidth]{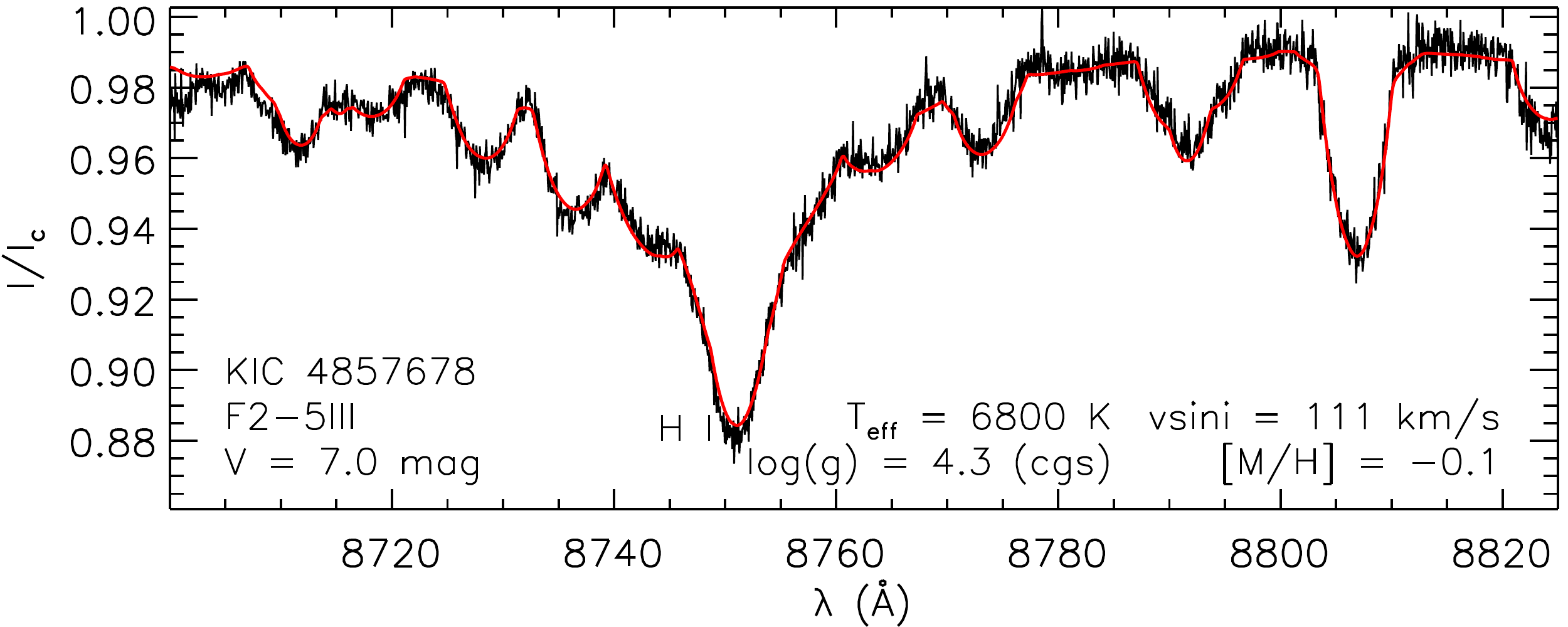}}\vspace{-0.4cm}
	\subfigure{\includegraphics[width=1\columnwidth]{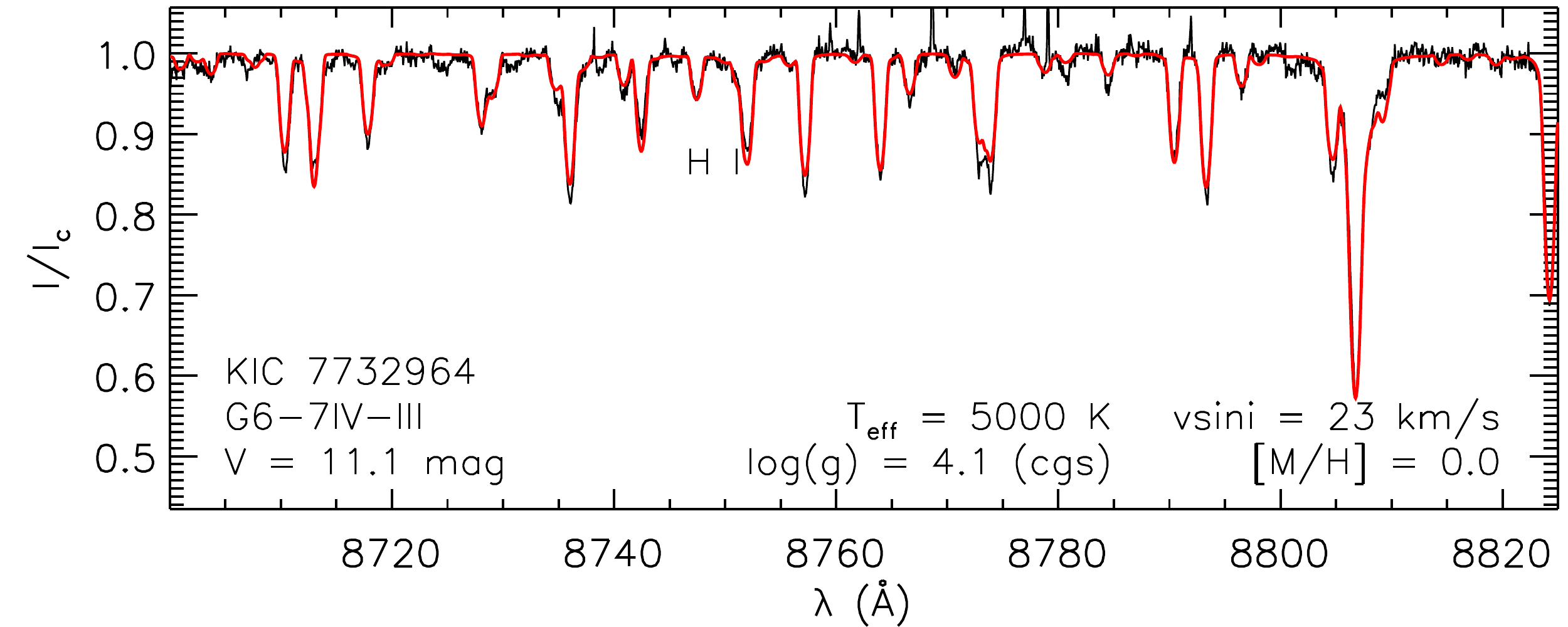}}\vspace{-0.4cm}
	\subfigure{\includegraphics[width=1\columnwidth]{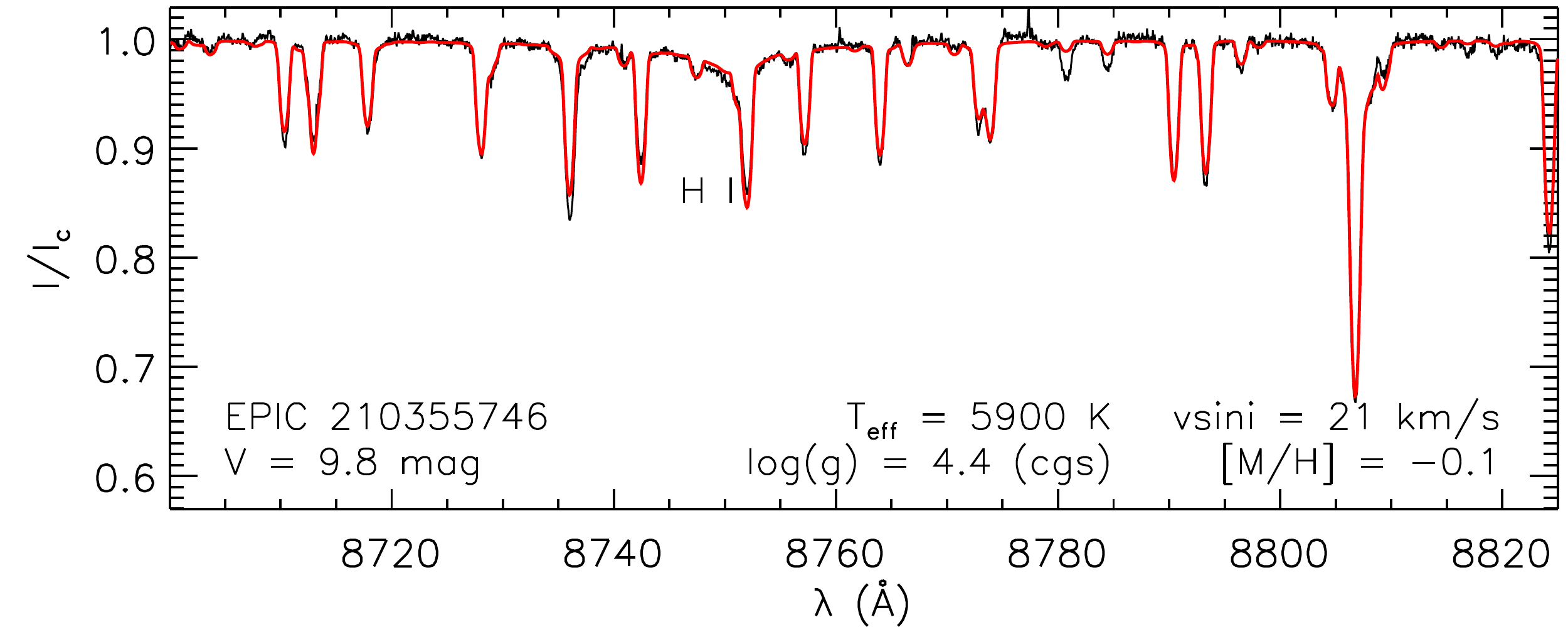}}\vspace{-0.4cm}
	\subfigure{\includegraphics[width=1\columnwidth]{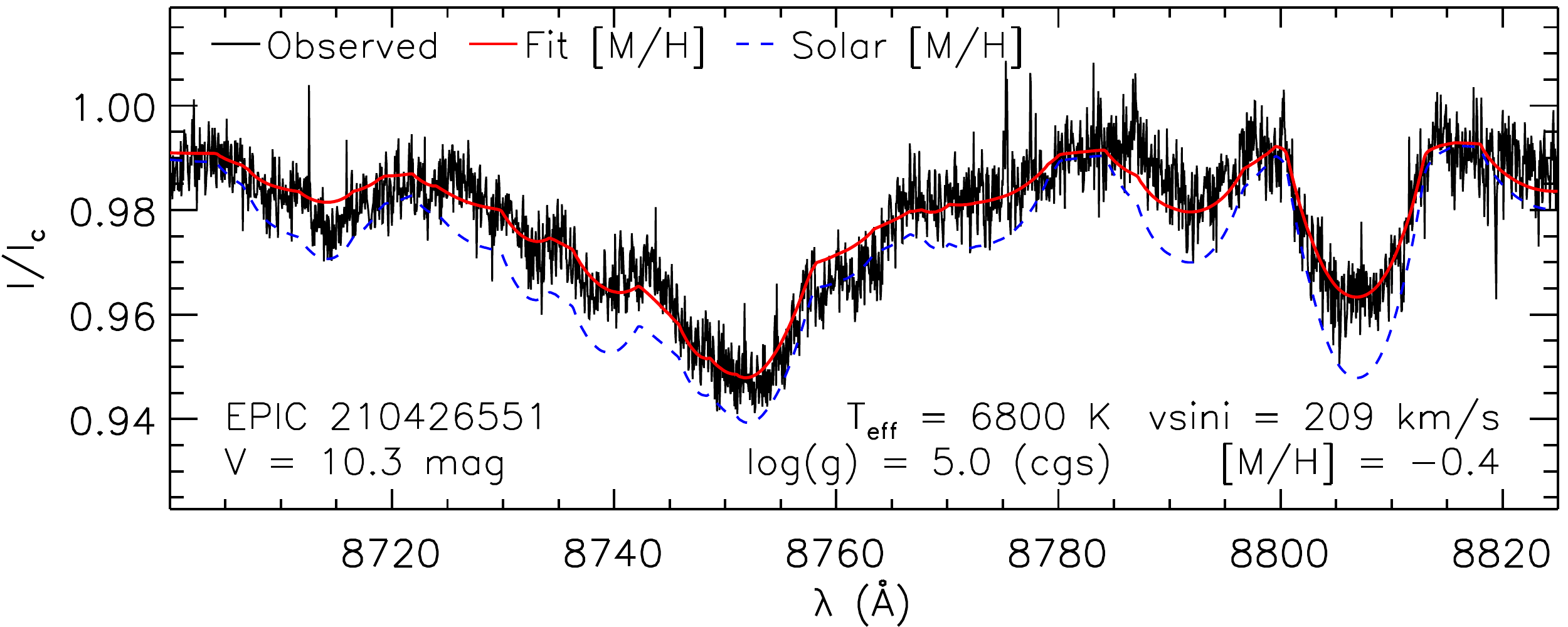}}\vspace{-0.4cm}
	\caption{Examples of the fits to the $8\,700-8\,825\,{\rm \AA}$ regions, which were used to derive $T_{\rm eff}$, $\log{g}$, [M/H], $v\sin{i}$, and $\xi$. The Paschen line (H{\sc i}~$\lambda8750$) centered in these regions is labeled. The black curve corresponds to the observed spectra and the solid red curve corresponds to the best-fitting model. For EPIC~210426551, we also plot the model generated using the same fitting parameters but with a solar metallicity ([M/H]$=0$) (dashed blue).}
	\label{fig:metal_fit}
\end{figure}

\begin{table*}
	\caption{Fundamental parameters derived from the spectral modelling analysis. Columns 1 and 2 list each target's ID and $T_{\rm eff}$ values taken from the \emph{Gaia} DR2 \citep{andrae2018}. Columns 3 to 7 list the derived $T_{\rm eff}$, $\log{g}$, $\xi$, $v\sin{i}$, and [M/H] values. All of the uncertainties associated with the derived parameters correspond to $1\sigma$.}
	\label{tbl:spec_param}
	\begin{center}
	\begin{tabular}{@{\extracolsep{\fill}}l c c c c c r@{\extracolsep{\fill}}}
		\hline
		\hline
		\noalign{\vskip0.5mm}
				ID  & $T_{\rm eff}^{\rm GDR2}$ & $T_{\rm eff}$ & $\log{g}$ & $\xi$          & $v\sin{i}$     & [M/H] \\
				    & (K)                      & (K)           & (cgs)     & (km$\,s^{-1}$) & (km$\,s^{-1}$) &       \\
				(1) & (2)                      & (3)           & (4)       & (5)            & (6)            & (7)   \\
		\noalign{\vskip0.5mm}
		\hline
		\noalign{\vskip0.5mm}\vspace{0.12cm}
EPIC 210355746  &  $5726_{-152}^{+209}$ &  $5900_{-140}^{+160}$ &           $4.4\pm0.2$ &   $1.0_{-0.3}^{+0.6}$ &                 $21\pm1$ &             $-0.1\pm0.1$ \\ \vspace{0.12cm}
EPIC 210384590  &   $5581_{-52}^{+119}$ &  $5500_{-120}^{+170}$ &   $4.4_{-0.2}^{+0.3}$ &   $1.5_{-0.6}^{+0.2}$ &                  $9\pm1$ &              $0.0\pm0.1$ \\ \vspace{0.12cm}
EPIC 210426551  &  $5802_{-179}^{+765}$ &  $6800_{-290}^{+100}$ &                $>4.6$ &                $<1.7$ &               $209\pm13$ &             $-0.4\pm0.1$ \\ \vspace{0.12cm}
EPIC 220533366  &  $4637_{-105}^{+131}$ &   $4700_{-70}^{+110}$ &   $3.2_{-0.2}^{+0.3}$ &   $1.5_{-0.3}^{+0.2}$ &                 $17\pm1$ &              $0.0\pm0.1$ \\ \vspace{0.12cm}
KIC 4857678     &  $6668_{-150}^{+106}$ &  $6800_{-170}^{+130}$ &   $4.3_{-0.3}^{+0.1}$ &                $<1.2$ &          $111_{-1}^{+6}$ &             $-0.1\pm0.1$ \\ \vspace{0.12cm}
KIC 6150124     &    $5055_{-62}^{+56}$ &   $5000_{-40}^{+240}$ &   $3.0_{-0.4}^{+0.5}$ &   $1.0_{-0.1}^{+0.6}$ &                  $4\pm1$ &             $-0.2\pm0.1$ \\ \vspace{0.12cm}
KIC 7732964     &   $5000_{-75}^{+114}$ &  $5000_{-200}^{+130}$ &           $4.1\pm0.4$ &   $1.5_{-0.6}^{+0.5}$ &           $23_{-2}^{+1}$ &              $0.0\pm0.1$ \\
		\hline \\
	\end{tabular}
	\end{center}
\end{table*}

Three regions within wavelengths of $7\,445$ and $8\,825\,{\rm \AA}$ and spanning approximately $100\,{\rm \AA}$ in width were selected to be modelled. The selected regions are free from tellurics and the density of spectral lines is lower than that of the shorter wavelength regions contained within the ESPaDOnS spectra thereby reducing line blends. The modelling that was carried out in order to derive $T_{\rm eff}$, $\log{g}$, [M/H], $v\sin{i}$, and $\xi$ proceeded using two sets of grid parameters: wide, low-resolution grids were used in the first iteration and narrower, higher-resolution grids were used in the second iteration. The high-resolution grids were used to derive the best-fitting parameters while the $\chi^2$ distributions associated with the low-resolution grids were used to estimate $1\sigma$ uncertainties for each parameter. The initial parameter grids were defined using the \emph{Gaia} DR2 $T_{\rm eff}$ values, a $\log{g}$ of 3 (cgs), a solar metallicity \citep{asplund2009}, a $\xi$ of $2\,{\rm km\,s^{-1}}$, and $v\sin{i}$ values that were estimated by eye. This procedure is similar to that employed by Sikora et al. (submitted). Ultimately, we found that the $8\,700-8\,825\,{\rm \AA}$ region, which contains the Paschen line corresponding to the $n=12$ to $n=3$ transition (H{\sc i}~$\lambda8750$), was more sensitive to changes in $T_{\rm eff}$ and $\log{g}$ compared to the $7\,450-7\,550\,{\rm \AA}$ and $7\,710-7\,810\,{\rm \AA}$ regions and yielded the highest-quality fits; therefore, the best-fitting parameters and their associated uncertainties were taken only from the fits to the $8\,700-8\,825\,{\rm \AA}$ regions.

We roughly evaluated the sensitivity of the $8\,700-8\,825\,{\rm \AA}$ region to changes in $\log{g}$ by using model spectra to estimate the maximum value of $\partial(I/I_{\rm c})/\partial(\log{g})$. For each star, we generated a grid of models spanning a range of $\log{g}$ in increments of $0.5$ while all other parameters were fixed at the best-fitting values. We found that for the $8\,700-8\,825\,{\rm \AA}$ region, $\partial(I/I_{\rm c})/\partial(\log{g})|_{\rm max}\sim0.02-0.1$; for comparison, we also computed $\partial(I/I_{\rm c})/\partial(\log{g})|_{\rm max}$ for a region centered on H$\alpha$ ($6\,530-6\,590\,{\rm \AA}$), which yielded comparable values. Therefore, we conclude that the adopted $8\,700-8\,825\,{\rm \AA}$ region is suitable for estimating $\log{g}$ to a degree that is acceptable within the scope of this study. We also note that the best-fitting $T_{\rm eff}$ values yielded by the fits to all three regions ($7\,450-7\,550$, $7\,710-7\,810$, and $8\,700-8\,825\,{\rm \AA}$) are in agreement within $500\,{\rm K}$.

Essentially all of the fitting parameters that were derived from the $8\,700-8\,825\,{\rm \AA}$ regions were reasonably well constrained; however, the $v\sin{i}$ value of EPIC~210426551 exhibited large $1\sigma$ uncertainties $\gtrsim30$~per~cent likely as a result of the high $v\sin{i}\gtrsim150\,{\rm km\,s^{-1}}$ and the relatively low S/N$\sim280$ (compared to the other spectra obtained for this study). In this case, we constrained $v\sin{i}$ by fitting several (apparently) unblended Fe and Mg lines within $7\,445-8\,825\,{\rm \AA}$ using the procedure described above. We adopted a $v\sin{i}$ value and associated uncertainty by taking the average and standard deviation of the resulting distribution. The values and the uncertainties of $T_{\rm eff}$, $\log{g}$, [M/H], and $\xi$ were then derived using the same method that was applied to the other stars while fixing $v\sin{i}$.

The $v\sin{i}$ values associated with the two components in the SB2 system, EPIC~210505125, were estimated using the Fast Fourier Transform method \citep[e.g.][]{gray_2005,simon-diaz2006}. This was carried out using various metal lines (e.g. Fe~{\sc i}, Mn~{\sc i}, Si~{\sc i}, and Cr~{\sc i}) located within the three $\approx100\,{\rm \AA}$ regions noted above. Each spectral component's adopted $v\sin{i}$ value and its uncertainty were taken to be the mean and standard deviation, respectively, of the resulting distribution. We obtained approximately equal values for the `A' and `B' components (as labeled in Table \ref{tbl:obs_vr_Bz}) of $12.8\pm1.8$ and $11.7\pm2.2\,{\rm km\,s}^{-1}$, respectively.

In Fig. \ref{fig:metal_fit}, we show several examples of the fits to the $8\,700-8\,825\,{\rm \AA}$ regions. Additional figures comparing the best-fitting model spectra (i.e. the model spectra generated using the adopted fitting parameters) with the observed spectra for three selected spectral windows ($4\,960-5050\,{\rm \AA}$, $7\,710-7\,810\,{\rm \AA}$, and $8\,700-8\,825\,{\rm \AA}$) are shown in Figures \ref{fig:spec_fit01} to \ref{fig:spec_fit07} in the online version of this paper. In Table \ref{tbl:spec_param}, we list the fundamental parameters ($T_{\rm eff}$, $\log{g}$, [M/H], $v\sin{i}$, and $\xi$) derived from the spectra. The table also lists the $T_{\rm eff}$ values found in the \emph{Gaia} DR2 catalog ($T_{\rm eff}^{\rm GDR2}$) for comparison; we find that both the derived $T_{\rm eff}$ values and the $T_{\rm eff}^{\rm GDR2}$ values are in agreement within their respective $1\sigma$ uncertainties for all of the stars. Note that the spectral modelling analysis was not carried out for the SB2 system EPIC~210505125.

\section{Hertzsprung-Russell Diagram}\label{sect:hrd}

As noted in Sect. \ref{sect:spec}, the $T_{\rm eff}$ values available in the \emph{Gaia} DR2 catalog \citep{andrae2018} are generally in agreement with the values derived from the spectral modelling analysis. However, in the case of EPIC~210426551, the difference between these two values ($\approx1\,000{\rm K}$) is significant. Spectroscopically-derived $T_{\rm eff}$ values are generally more accurate than those derived using photometric calibrations \citep[particularly for stars with $T_{\rm eff}\gtrsim7\,000{\rm K}$, e.g.][]{molenda-zakowicz2010,brown2011}; therefore, we opted to place our sample of stars on the Hertzsprung-Russell diagram (HRD) using the $T_{\rm eff}$ values derived from the spectra in conjunction with luminosities derived using the distance estimates that have been inferred from the \emph{Gaia} DR2 parallax measurements \citep{gaiacollaboration2018}.

\begin{table*}
	\caption{Various parameters associated with the HRD. Columns 2 to 4 list the distance reported by \citet{bailer-jones2018} based on \emph{Gaia} DR2 parallax measurements and the derived or adopted $R$ and $\log{L}$ values. Columns 4 to 7 list the masses ($M$), ages ($t_{\rm age}$), fractional MS ages ($\tau$), and critical rotational velocities ($v_{\rm crit}$).}
	\label{tbl:hrd_param}
	\begin{center}
	\begin{tabular}{@{\extracolsep{\fill}}l c c c c c c r@{\extracolsep{\fill}}}
		\hline
		\hline
		\noalign{\vskip0.5mm}
				ID  & $d$  & $R$         & $\log{(L/L_\odot)}$ & $M$         & $t_{\rm age}$ & $\tau$ & $v_{\rm crit}$       \\
				    & (pc) & ($R_\odot$) &                     & ($M_\odot$) & (Gyrs)        &        & (${\rm km\,s}^{-1}$) \\
				(1) & (2)  & (3)         & (4)                 & (5)         & (6)           & (7)    & (8)                  \\
		\noalign{\vskip0.5mm}
		\hline
		\noalign{\vskip0.5mm}\vspace{0.12cm}
EPIC 210355746  &  $125.2\pm0.7$ &                    $1.18\pm0.04$ &                 $0.18\pm0.07$ &        $0.97_{-0.04}^{+0.07}$ &        $9.99_{-0.24}^{+0.08}$ &        $1.10_{-0.30}^{+0.09}$ &             $307_{-11}^{+29}$ \\ \vspace{0.12cm}
EPIC 210384590  &   $61.2\pm0.2$ &                    $0.94\pm0.04$ &                $-0.14\pm0.08$ &        $0.86_{-0.03}^{+0.08}$ &          $10.1_{-0.4}^{+0.2}$ &           $0.9_{-0.4}^{+0.2}$ &                    $336\pm19$ \\ \vspace{0.12cm}
EPIC 210426551  &      $184\pm2$ &                    $1.26\pm0.09$ &                   $0.5\pm0.1$ &        $1.05_{-0.05}^{+0.09}$ &                   $9.6\pm0.3$ &           $1.1_{-0.4}^{+0.1}$ &             $306_{-16}^{+37}$ \\ \vspace{0.12cm}
EPIC 210505125  &   $77.0\pm0.3$ & ${1.06_{-0.07}^{+0.06}}^\dagger$ &     ${0.134\pm0.002}^\dagger$ &        $1.04_{-0.07}^{+0.08}$ &           $9.8_{-0.5}^{+0.2}$ &           $0.8_{-0.4}^{+0.2}$ &                    $343\pm16$ \\ \vspace{0.12cm}
EPIC 220533366  &      $254\pm4$ &                      $7.1\pm0.4$ &                 $1.34\pm0.08$ &           $1.2_{-0.1}^{+0.4}$ &           $9.4_{-0.2}^{+0.4}$ &                   $1.3\pm0.2$ &                    $154\pm13$ \\ \vspace{0.12cm}
KIC 4857678     & $54.66\pm0.08$ &                    $1.36\pm0.04$ &                 $0.55\pm0.06$ &        $1.29_{-0.07}^{+0.05}$ &           $9.4_{-0.3}^{+0.1}$ &           $0.7_{-0.3}^{+0.2}$ &                    $331\pm18$ \\ \vspace{0.12cm}
KIC 6150124     &      $257\pm2$ &                     $12.2\pm0.5$ &                 $1.92\pm0.09$ &                   $2.2\pm0.3$ &        $8.77_{-0.24}^{+0.06}$ &               $1.033\pm0.009$ &             $182_{-34}^{+18}$ \\ \vspace{0.12cm}
KIC 7732964     &      $182\pm4$ &                    $1.43\pm0.08$ &                   $0.1\pm0.1$ &        $0.77_{-0.03}^{+0.09}$ &                  $10.3\pm0.1$ &        $1.26_{-0.04}^{+0.03}$ &                    $263\pm12$ \\
		\noalign{\vskip0.5mm}
		\hline \\
\noalign{\vskip-0.3cm}
\multicolumn{8}{l}{$^\dagger$Taken from \citet{andrae2018} and adjusted to account for binarity ($L\rightarrow L/2$ and $R\rightarrow R/\sqrt{2}$).} \\
	\end{tabular}
	\end{center}
\end{table*}

We derived radii ($R$) and luminosities ($L$) for the 7 stars with spectroscopically-derived $T_{\rm eff}$ values by fitting synthetic spectral energy distributions (SEDs) to published photometric measurements. This was carried out using a procedure similar to that discussed by Sikora et al. (submitted). Measurements obtained using various photometric filters were compiled including Johnson $BV$, Tycho $B_TV_T$ \citep{esa1997}, Str\"omgren $ubvy$ \citep{hauck1998}, \emph{Gaia} DR2 $GG_{BP}G_{RP}$ \citep{gaiacollaboration2018}, 2MASS $JHK$ \citep{cohen2003}, and WISE $W_1W_2W_3W_4$ \citep{wright2010}. We used reddening parameters (E[$B-V$]) published by \citet{gontcharov2017} and distances were taken from the catalog published by \citet{bailer-jones2018} based on \emph{Gaia} DR2 parallax measurements. Synthetic SEDs published by \citet{castelli2003} were fit to the photometric measurements by fixing $T_{\rm eff}$, $\log{g}$, and [M/H] to the best-fitting values listed in Table \ref{tbl:spec_param} while allowing $R$ to vary. The uncertainty in $R$ was estimated using a bootstrapping procedure and subsequently including the uncertainty in the distances, as described by Sikora et al. (submitted). Finally, luminosities were derived using the $T_{\rm eff}$ and $R$ values through the Stefan-Boltzmann relation.

EPIC~210505125 is an SB2 system that is unresolved within the \emph{Gaia} DR2 catalog \citep{gaiacollaboration2018}. The two stars have similar spectral types (see Fig. \ref{fig:all_lsd}) and therefore their reported $T_{\rm eff}$ values are unlikely to be significantly altered from their true values. The luminosities of the individual stars, on the other hand, are likely overestimated by approximately a factor of 2. We adopt the effective temperature and half of the luminosity ($L/2$) reported within the \emph{Gaia} DR2 catalog \citep{andrae2018}.

\begin{figure}
	\centering
	\includegraphics[width=1.0\columnwidth]{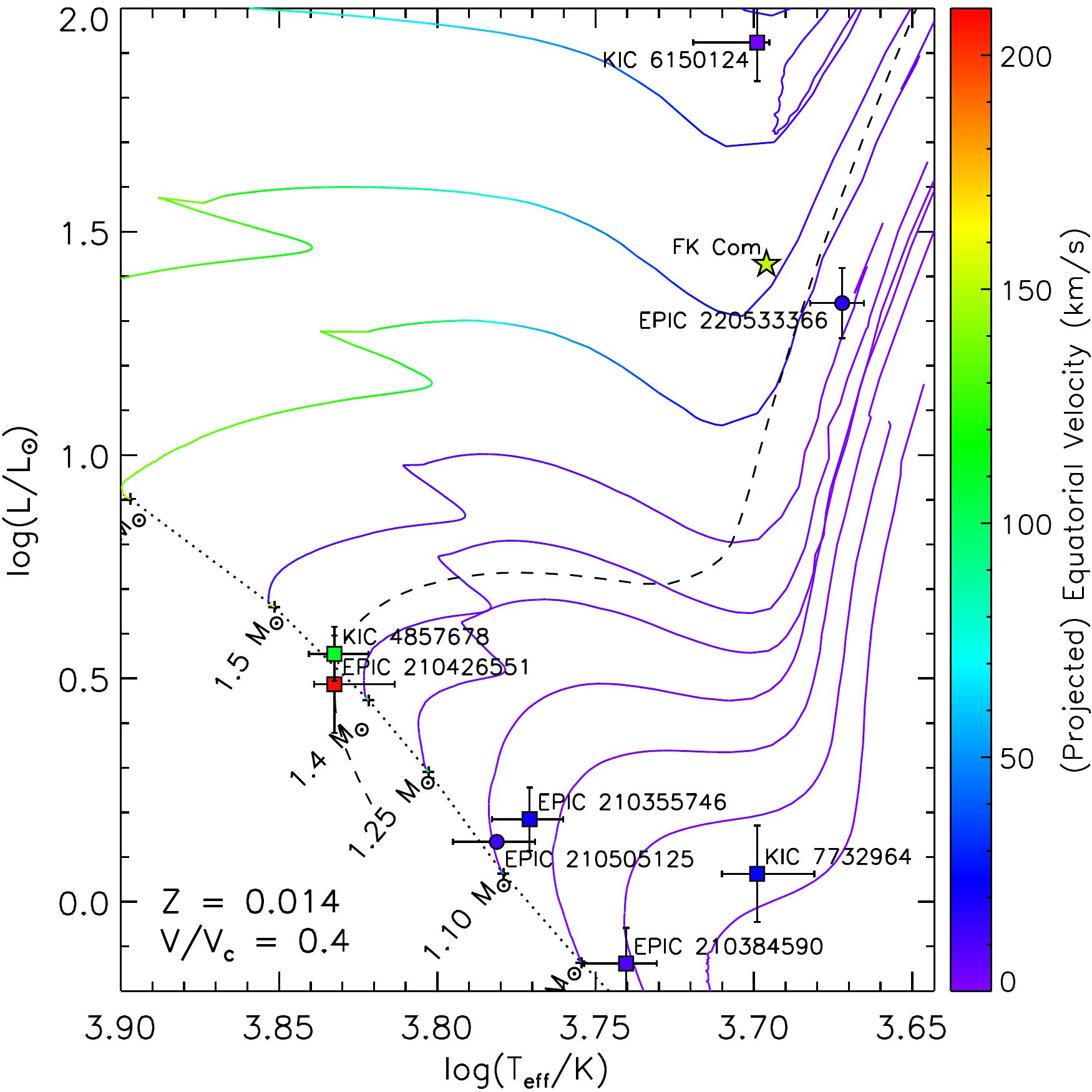}\vspace{-0.15cm}
	\caption{HRD associated with the targets observed for this study compared with the solar metallicity ($Z=0.014$), rotating models ($v_{\rm eq}/v_{\rm c}=0.4$) generated by \citet{ekstrom2012} (solid lines); the dotted black line indicates the ZAMS. Circle symbols correspond to spectroscopic binaries while square symbols correspond to those stars for which no spectrscopic signatures of companions were detected. FK~Com is plotted for reference based on its \emph{Gaia} DR2 stellar parameters \citep{andrae2018}. The colours of the square and circle symbols correspond to the $v\sin{i}$ values of those objects while the colours of the evolutionary tracks correspond to the associated equatorial velocities. We also plot the best-fitting non-rotating evolutionary track associated with EPIC~210426551, which corresponds to $M=1.08\,M_\odot$ and $Z=0.006$ (${\rm [M/H]}=-0.4$) (dashed black).}
	\label{fig:HRD}
\end{figure}

In Fig. \ref{fig:HRD}, we show the HRD containing the 8 targets in our sample colour-coded by their derived $v\sin{i}$ values. We also plot the grid of solar-metallicity, rotating ($v/v_{\rm crit}=0.4$) evolutionary model tracks published by \citet{ekstrom2012}. The non-rotating $Z=0.002$ and $Z=0.014$ grids of evolutionary models published by \citet{georgy2013} and \citet{ekstrom2012} along with the non-rotating $Z=[0.006,0.04]$ grids published by \citet{mowlavi2012} were linearly interpolated in order to derive each star's mass and age (i.e. evolutionary state) based on the adopted $T_{\rm eff}$, $L$, and [M/H] values. For EPIC~210505125, which is located at a distance of $77.0\pm0.3\,{\rm pc}$,we assume ${\rm [M/H]}=0.0\pm0.2$ where the uncertainty corresponds to the standard deviation of the solar neighbourhood's ${\rm [M/H]}$ distribution reported by \citet{casagrande2011}.

Uncertainties in the parameters derived from the evolutionary models were estimated using the Monte Carlo method discussed by \cite{sikora2019}, which is based on that used by \citet{shultz2019a}. This involves generating $\sim3\,000$ $T_{\rm eff}$, $L$, and [M/H] points randomly sampled from distributions defined by the adopted values and their uncertainties. For each of these points, relevant parameters are derived from the evolutionary grids (mass, age, etc.) and the resulting distributions are used to estimate 1$\sigma$ uncertainties. Note that in several cases, the randomly sampled distributions contain [M/H] values that fall outside of the model grid limits; therefore, in these cases the upper uncertainty in mass and lower uncertainty in age may be truncated. Furthermore, the models do not include the pre-MS, which also contributes to the truncation of the lower age uncertainty. In Table \ref{tbl:hrd_param}, we list the derived (and adopted) radii and luminosities along with the derived masses and ages for the 8 stars in our sample. We also report the fractional MS age (i.e. the age divided by the MS lifetime, $\tau$) and the critical equatorial rotational velocity \citep[$v_{\rm crit}$, as defined in Eqn. 3.14 of][]{maeder2000}.

Most of the stars in our sample that were included in the spectral modelling analysis (Sect. \ref{sect:spec}) were found to have metallicities approximately consistent with solar values \citep{asplund2009}. The largest deviation from a solar metallicity is exhibited by EPIC~210426551 \citep[${\rm [M/H]}=-0.4\pm0.1$, which is within $2.4\sigma$ of the average metallicity within the solar neighbourhood;][]{casagrande2011}. Comparing the grids of evolutionary models, it is clear that adopting a lower metallicity will yield a lower inferred mass and imply that the star is more evolved. Therefore, in Fig. \ref{fig:HRD}, we plot the best-fitting evolutionary track associated with EPIC~210426551 (i.e. the interpolated model that passes through the derived $T_{\rm eff}$, $L$, and [M/H] values).

\begin{table}
	\caption{Parameters associated with or derived from the \emph{Kepler} and \emph{K2} light curves. Column 2 lists the minimum rotation period estimated using the $v_{\rm crit}$ values listed in Table \ref{tbl:hrd_param}. Column 3 lists the minimum rotation period estimated from the derived $v\sin{i}$ values. Column 4 lists the periods identified in the light curves that are most likely attributed to the star's rotation and column 5 lists the equatorial velocity inferred from $P_{\rm phot}$ (under the assumption that $P_{\rm phot}=P_{\rm rot}$). Values listed in parentheses correspond to uncertainties in the last digit.}
	\label{tbl:LC}
	\begin{center}
	\begin{tabular}{@{\extracolsep{\fill}}l c c c r@{\extracolsep{\fill}}}
		\hline
		\hline
		\noalign{\vskip0.5mm}\vspace{0.12cm}
				ID  & $P_{\rm rot}^{\rm min}$ & $P_{\rm rot}^{\rm max}$ & $P_{\rm phot}$ & $v_{\rm eq}^{\rm phot}$ \\ \vspace{0.12cm}
				    & $({\rm d})$             & $({\rm d})$             & $({\rm d})$    & $({\rm km\,s}^{-1})$    \\
				(1) & (2)                     & (3)                     & (4)            & (5)                     \\
		\noalign{\vskip0.5mm}
		\hline
		\noalign{\vskip0.5mm}
EPIC 210355746  &  0.18 &   3.1 &    2.929717(4) & 20.5(6) \\
EPIC 210384590  &  0.14 &   6.0 &    5.152970(4) & 9.22(4) \\
EPIC 210426551  &  0.20 &  0.35 & 0.346259240(3) & 184(12) \\
EPIC 210505125  &  0.16 &   6.0 &     3.81298(1) & 14.1(8) \\
EPIC 220533366  &   2.2 &  22.9 &    14.00744(1) &   26(1) \\
KIC 4857678     &  0.20 & 0.641 &                &         \\
KIC 6150124     &   3.0 &   200 &     31.3698(1) & 19.7(8) \\
KIC 7732964     &  0.26 &   3.7 &   2.4441471(1) &   30(2) \\
		\noalign{\vskip0.5mm}
		\hline \\
	\end{tabular}
	\end{center}
\end{table}

\section{Light Curves}\label{sect:LC}

\begin{figure*}
	\centering
	\begin{minipage}{1.0\columnwidth}
		\subfigure{\includegraphics[width=1\columnwidth]{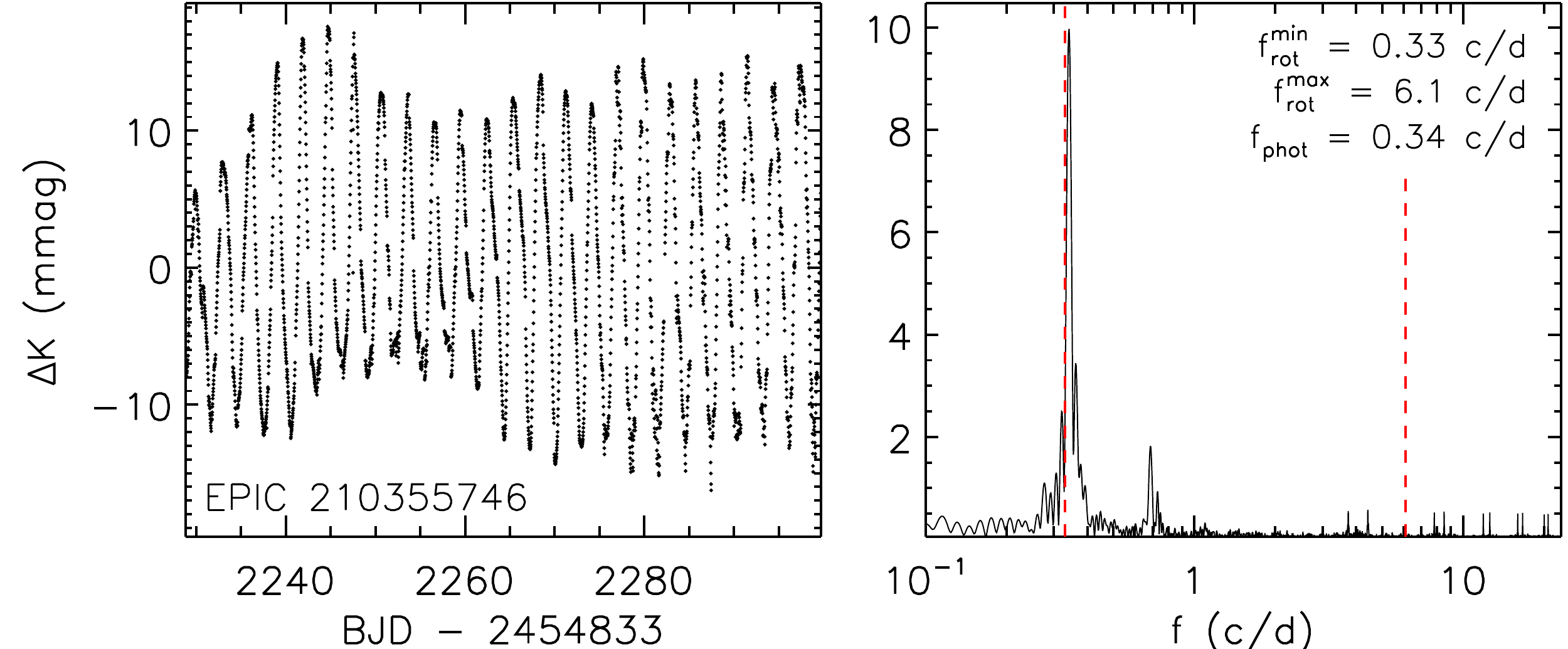}}\vspace{-0.36cm}
		\subfigure{\includegraphics[width=1\columnwidth]{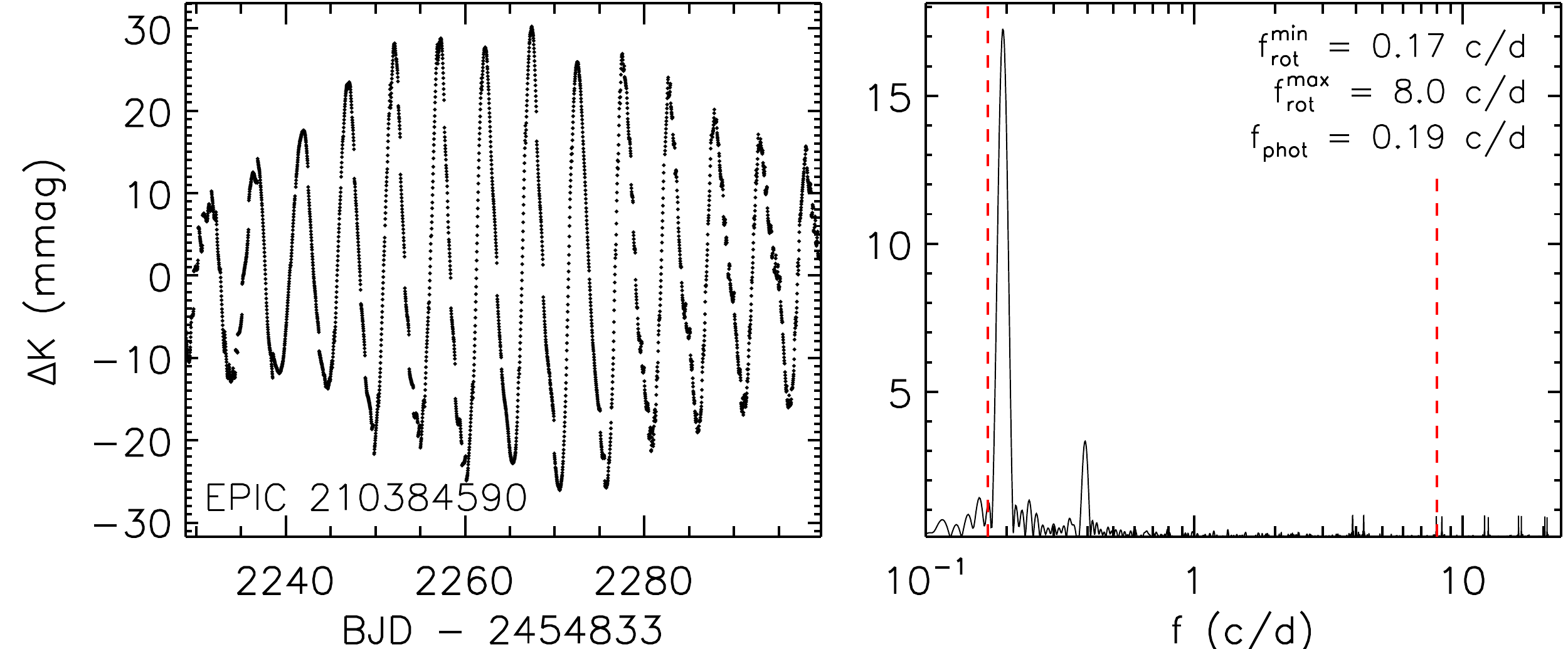}}\vspace{-0.36cm}
		\subfigure{\includegraphics[width=1\columnwidth]{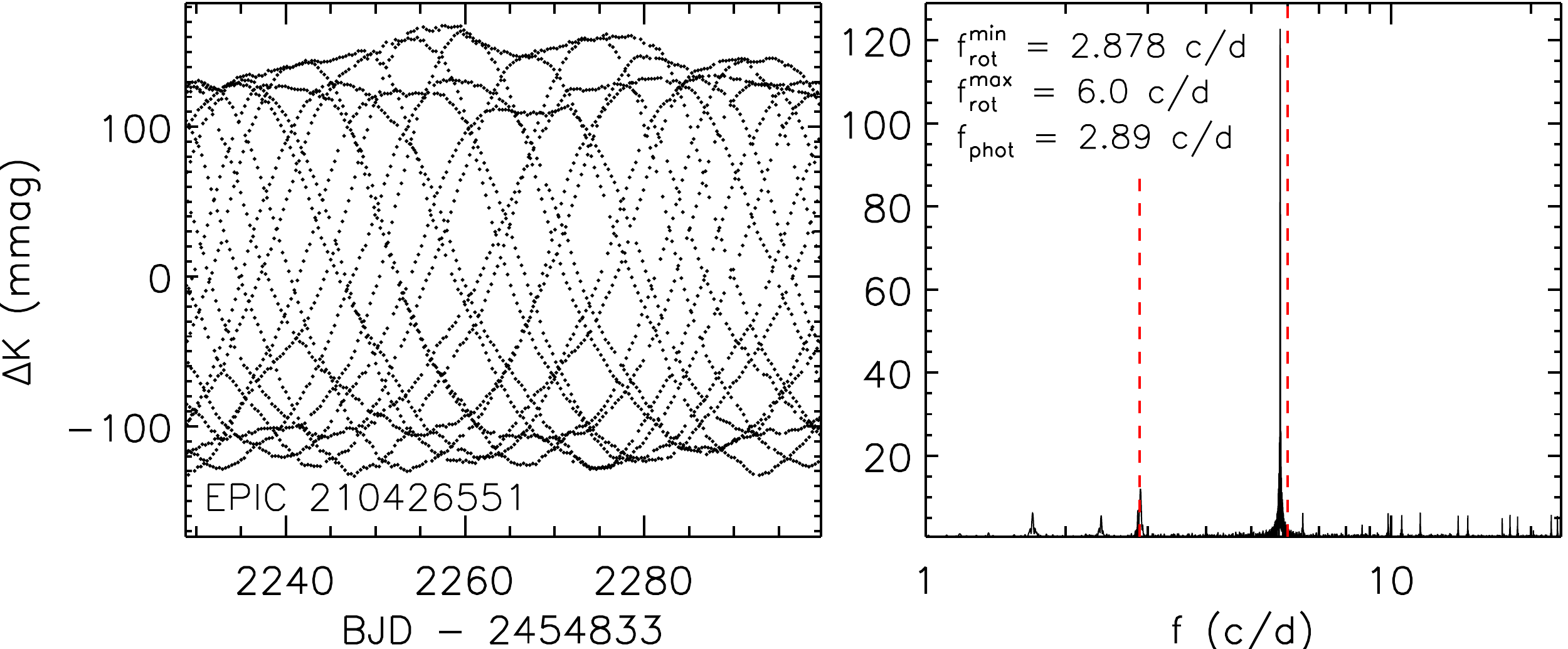}}\vspace{-0.36cm}
		\subfigure{\includegraphics[width=1\columnwidth]{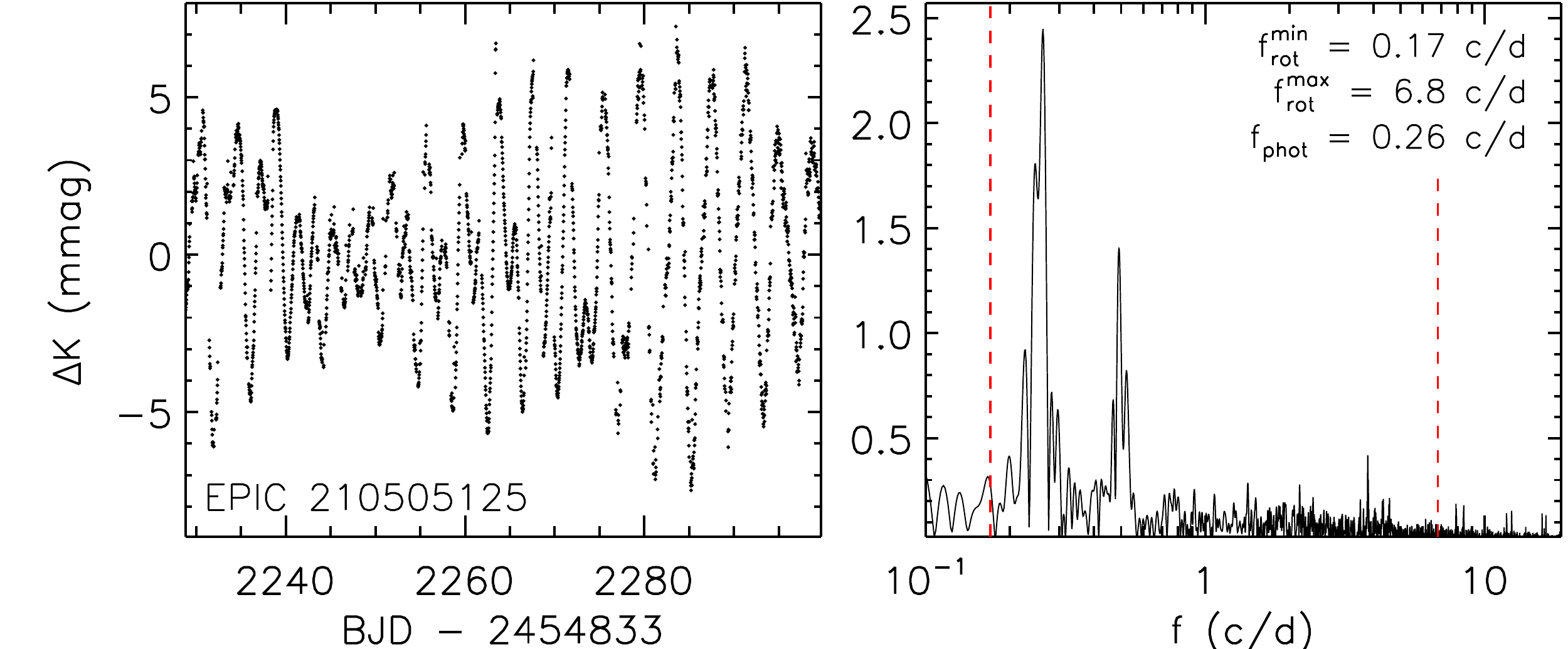}}
	\end{minipage}
	\begin{minipage}{1.0\columnwidth}
		\subfigure{\includegraphics[width=1\columnwidth]{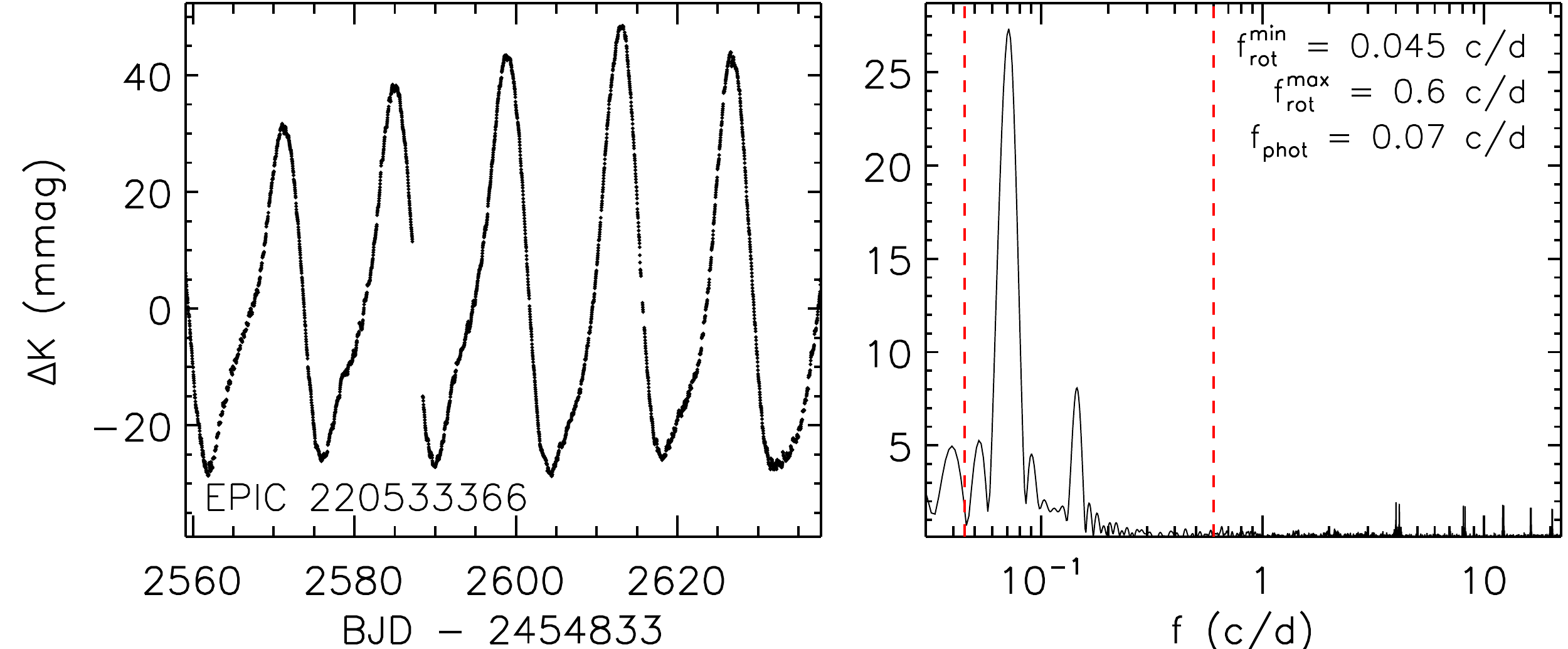}}\vspace{-0.36cm}
		\subfigure{\includegraphics[width=1\columnwidth]{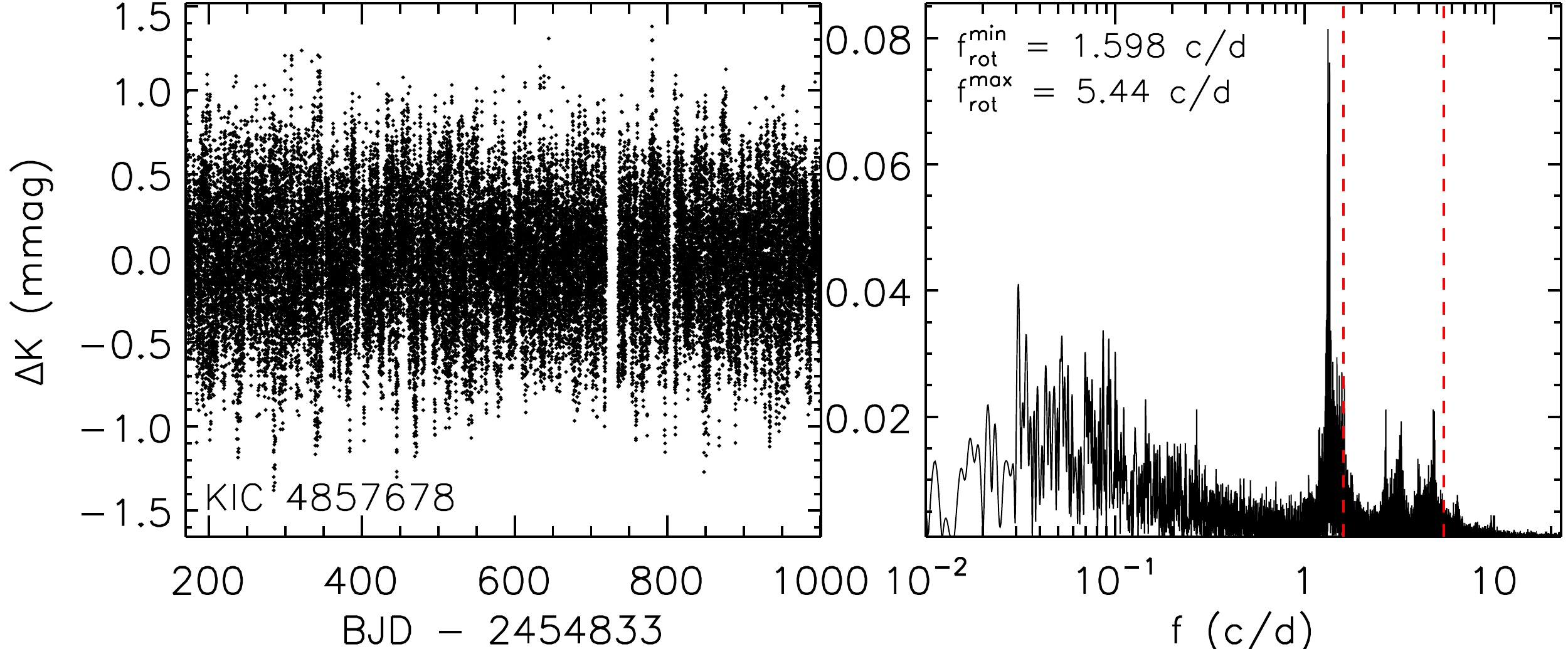}}\vspace{-0.36cm}
		\subfigure{\includegraphics[width=1\columnwidth]{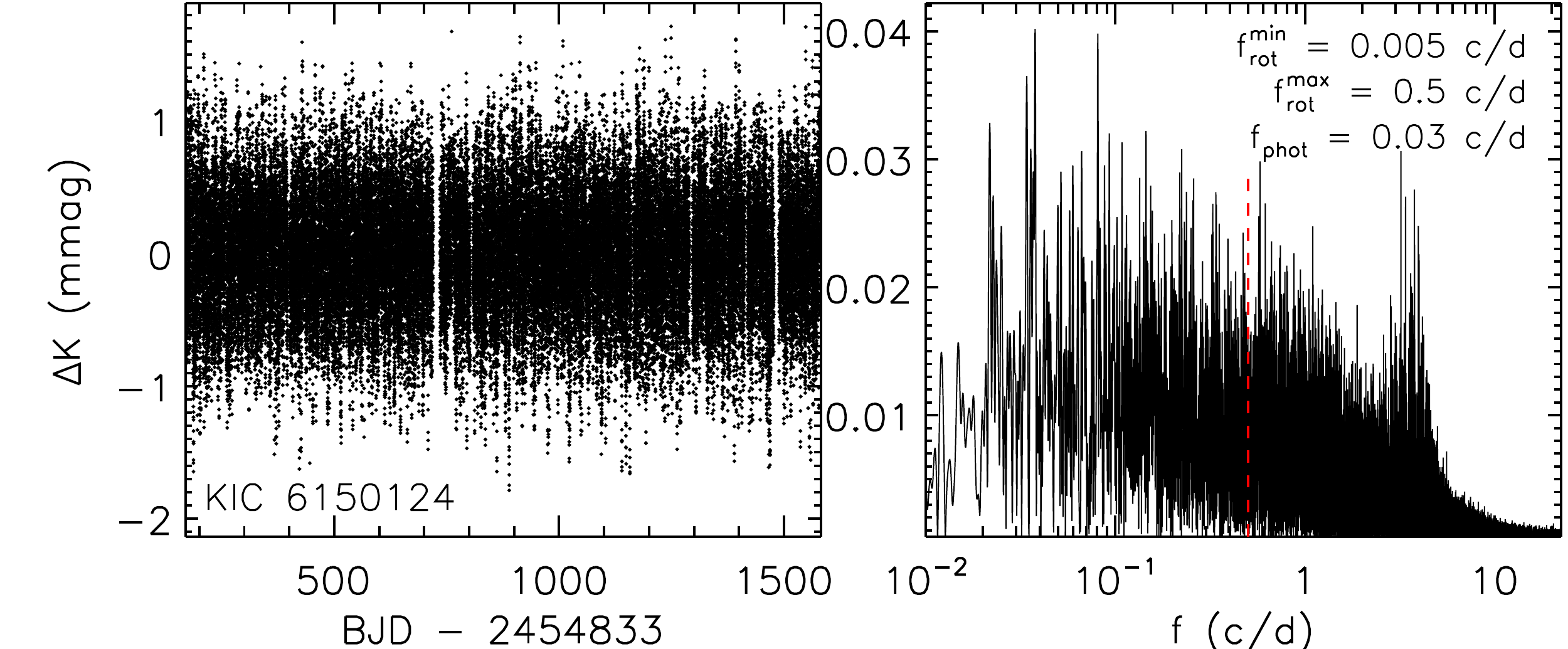}}\vspace{-0.36cm}
		\subfigure{\includegraphics[width=1\columnwidth]{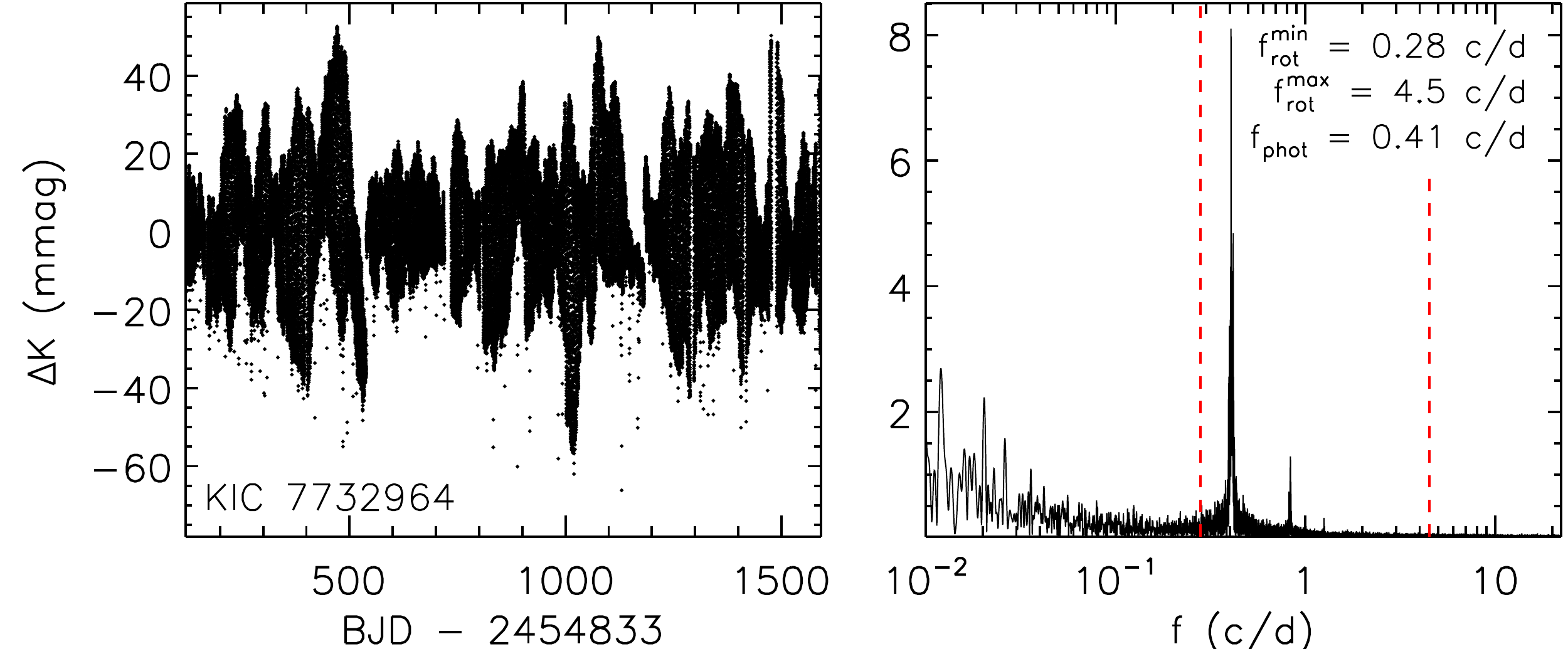}}
	\end{minipage}
	\caption{\emph{Kepler} and \emph{K2} light curves. The first and third columns show the full extracted and post-processed light curves; to the right of each light curves, we show the associated LS periodogram. For each of the periodograms, at least one peak is identified within the estimated lower and upper rotational frequency limits ($f_{\rm rot}^{\rm min}$ and $f_{\rm rot}^{\rm max}$, respectively), which are indicated by the red dashed lines. We note that these peaks may be associated with oscillation frequencies (e.g. KIC~6150124) or may correspond to integer multiples of $f_{\rm rot}$. The photometric periods that can plausibly be attributed to each star's rotation period ($P_{\rm phot}=1/f_{\rm phot}$) are listed in Table \ref{tbl:LC}.}
	\label{fig:LC}
\end{figure*}

The \emph{Kepler} and \emph{K2} light curves used in this study are shown in Fig. \ref{fig:LC} along with their associated Lomb-Scargle (LS) periodograms \citep{lomb1976,scargle1982}. All of the light curves exhibit clear variability with amplitudes ranging from $\sim0.1-100\,{\rm mmag}$. Multiple peaks with frequencies $\lesssim10\,{\rm d}^{-1}$ are apparent in each of the LS periodograms. While some of these signals correspond to oscillation frequencies \citep[e.g. $f\approx3.5\,{\rm d}^{-1}$ in KIC~6150124,][]{yu2018}, others may be associated with the presence of spots on the rotating star's surface and thus, they may correspond to the star's rotation period ($P_{\rm rot}$).

In order to identify potential rotational modulation signatures, we estimated lower and upper frequency limits based on (1) the $v\sin{i}$ values derived in Sect. \ref{sect:spec} ($f_{\rm rot}^{\rm min}=1/P_{\rm rot}^{\rm max}\equiv v\sin{i}/2\pi R$) and (2) the $v_{\rm crit}$ values derived in Sect. \ref{sect:hrd} ($f_{\rm rot}^{\rm max}=1/P_{\rm rot}^{\rm min}\equiv v_{\rm crit}/2\pi R$). These $P_{\rm rot}^{\rm min}$ and $P_{\rm rot}^{\rm max}$ values (listed in Table \ref{tbl:LC}) are computed using the lower and upper bounds of each relevant parameter. For each of the 8 targets, we searched for those peaks with frequencies within the $f_{\rm rot}^{\rm min}$ and $f_{\rm rot}^{\rm max}$ limits, which can plausibly be attributed to $P_{\rm rot}$. In those cases where multiple peaks are present, we generally adopted that with the highest amplitude (as discussed below, EPIC~210426551 and KIC~4857678 are two exceptions). The resulting photometric periods ($P_{\rm phot}$) were then refined by fitting a sinusoidal function consisting of $P_{\rm phot}$ and its first four harmonics \citep[see][]{sikora2019b}. In Table \ref{tbl:LC} we report the derived $P_{\rm phot}$ values along with their $1\sigma$ uncertainties. We also report the equatorial velocities derived under the assumption that $P_{\rm phot}=P_{\rm rot}$ ($v_{\rm eq}^{\rm phot}$).

For 5 of the targets (EPIC~210355746, EPIC~210384590, EPIC~210505125, EPIC~220533366, and KIC~7732964), the identified $f_{\rm phot}$ ($1/P_{\rm phot}$) values correspond to the maximum amplitude signals that appear in the full LS periodogram (i.e. below the Nyquist frequency). Furthermore, these $f_{\rm phot}$ values do not appear to be harmonics of lower frequencies and also exhibit a first harmonic. In these cases, it is very likely that $P_{\rm phot}$ is the rotation period. The 3 targets for which this correspondance between $P_{\rm phot}$ and $P_{\rm rot}$ is more ambiguous are discussed below.

For EPIC~210426551, the highest amplitude peak has a frequency of $5.776\,{\rm d}^{-1}$, however, it is likely the first harmonic of the lower-amplitude peak appearing at $2.888\,{\rm d}^{-1}$. This lower frequency falls slightly above $f_{\rm rot}^{\rm min}$ and we therefore adopt it as the $f_{\rm phot}$ value (i.e. that which is most likely to correspond to $f_{\rm rot}$). For KIC~4857678, the highest amplitude peak has a frequency of $1.329\,{\rm d}^{-1}$, which corresponds to the $0.75\,{\rm d}$ photometric period reported by \citet{howell2016}. This frequency falls below $f_{\rm rot}^{\rm min}=1.598\,{\rm d}^{-1}$ implying that it cannot correspond to $f_{\rm rot}$. The complex peak distribution of KIC~4857678's periodogram is characteristically similar to that observed for the pulsating $\gamma$~Dor stars \citep[e.g.][]{balona2011b,antoci2019}. Furthermore, KIC~4857678 has a similar spectral type (early to mid F) to the $\gamma$~Dor stars \citep[e.g.][]{krisciunas1993}. We conclude that the variability detected in the light curve is most likely pulsational in origin and do not report a $P_{\rm phot}$ value. Finally, for KIC~6150124 a large number of peaks are apparent within the $f_{\rm rot}$ limits and are thus consistent with the star's rotation frequency, however, these signals cannot easily be disentangled from the increased power at $f\lesssim1\,{\rm d}^{-1}$ caused by granulation \citep{yu2018}.

\section{Discussion}\label{sect:disc}

It has been suggested that the 8 targets (3 from the \emph{Kepler} mission and 5 from the \emph{K2} mission) that were included in this study are possible candidates of the FK~Comae type stars based on their enhanced X-ray and UV emission and their apparent rapid rotation rates \citep{howell2016}. As discussed below, we find that none of the 8 targets can be considered FK~Comae candidates based on the properties listed in Table \ref{tbl:fk_comae}.

The 3 \emph{Kepler} targets that were observed in this study were also observed using a medium-resolution spectrograph by \citet{howell2016}, who reported that all of these stars exhibit $70\lesssim v\sin{i}\lesssim140\,{\rm km\,s}^{-1}$. We derived $v\sin{i}$ values based on the newly obtained high-resolution spectra and found that, while these values are in agreement for KIC~4857678, our $v\sin{i}$ values are significantly lower for KIC~6150124 ($4\pm1\,{\rm km\,s}^{-1}$ compared to a reported minimum value of $72.8\pm14.6\,{\rm km\,s}^{-1}$) and KIC~7732964 ($23_{-2}^{+1}\,{\rm km\,s}^{-1}$ compared to a minimum value of $98.4\pm19.7\,{\rm km\,s}^{-1}$). The discrepancies can likely be attributed to the fact that our observations have higher S/Ns ($\approx200-600$ compared to $36-50$) and were obtained with a higher resolving power ($R\sim65\,000$ compared to $R\sim7\,700-10\,000$). Therefore, it is likely that both relatively high S/Ns ($\gtrsim100$) and high-resolution instruments are required in order to exclude false positives in the search for FK~Comae type stars.

\begin{figure*}
	\centering
	\begin{minipage}{0.9\columnwidth}
		\centering
		\includegraphics[width=1.0\columnwidth]{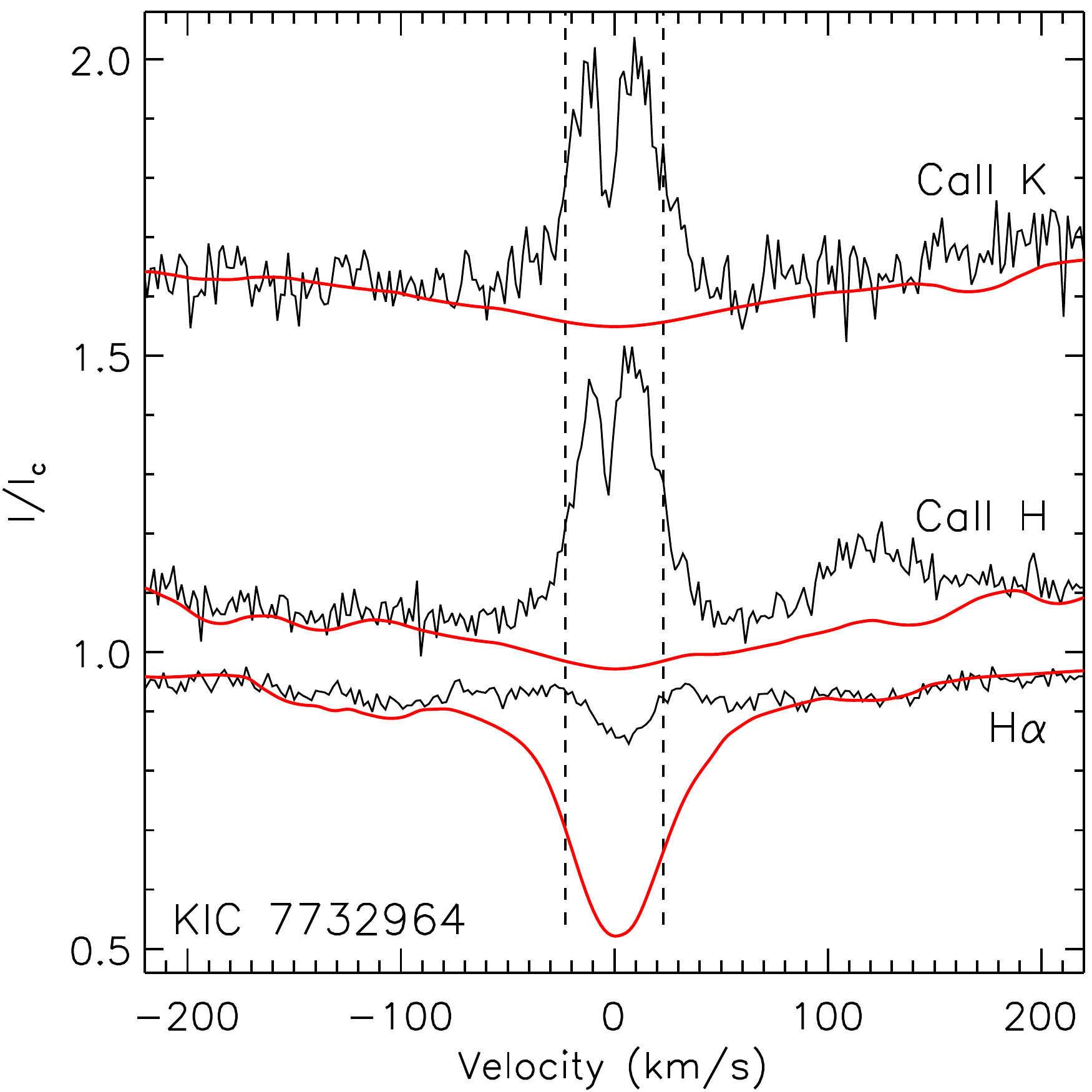}
	\end{minipage}
	\begin{minipage}{0.9\columnwidth}
		\centering
		\includegraphics[width=1.0\columnwidth]{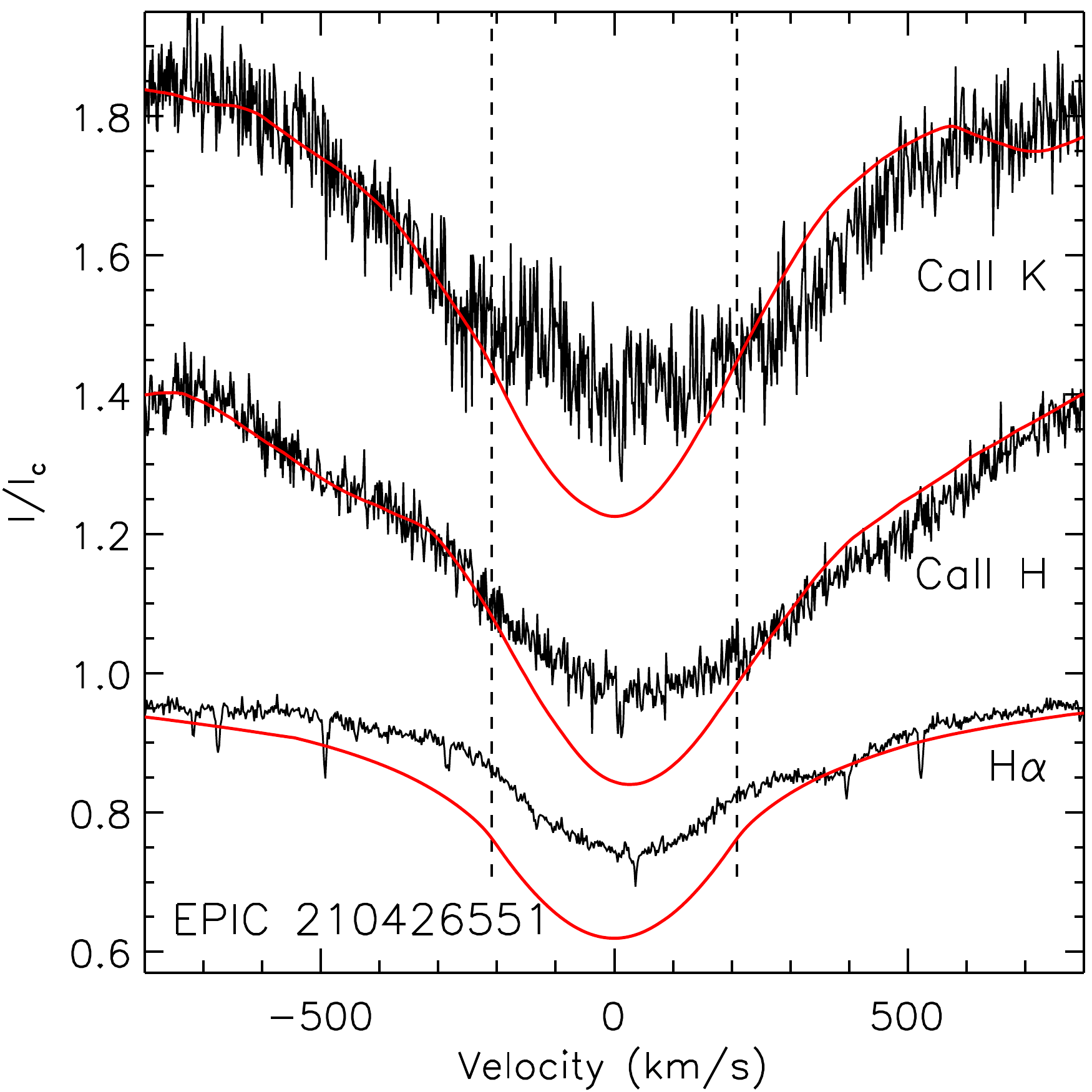}
	\end{minipage}
	\caption{Observed spectra (black curves) compared with the synthetic models (red curves) generated using the best-fitting fundamental parameters (Table \ref{tbl:spec_param}) for Ca~{\sc ii}~K, Ca~{\sc ii}~H, and H$\alpha$. The vertical dashed lines correspond to each star's $v\sin{i}$ values. \emph{Left:} Clear, double-peaked emission is seen in all three lines for KIC~7732964. \emph{Right:} For EPIC~210426551, we detect weak Ca~{\sc ii}~H and K emission. The apparent emission in H$\alpha$ is primarily inferred from the residuals between the observed and model profiles, however, it may be related to systematic errors in the normalization. In both cases, the detection of emission suggests that these stars are magnetically active.}
	\label{fig:emission}
\end{figure*}

Out of the 8 targets included in our study, 5 are found to have $v\sin{i}$ values ranging from $9-23\,{\rm km\,s}^{-1}$. The \emph{Kepler} and \emph{K2} light curves for these targets all reveal photometric variability that is consistent with rotational modulation. We derived rotational periods for these targets using the light curves, which, when combined with the stellar radii, yield equatorial velocities $\lesssim30\,{\rm km\,s}^{-1}$. The estimated rotation rates for these stars are considered anomalously high amongst evolved stars \citep[e.g.][]{drake2002,massarotti2008} but are not unusual amongst main sequence (MS) or pre-MS stars \citep[e.g.][]{skumanich1972,hartmann1986,mcquillan2014}. Three of the 5 stars in this $v\sin{i}$ range (EPIC~210505125, EPIC~210355746, and EPIC~210384590) lie along the MS, which is consistent with their derived $\log{g}$ values of $\approx4.4\,{\rm (cgs)}$. EPIC~220533366 is a red giant branch (RGB) star with $v\sin{i}=17\pm1\,{\rm km\,s}^{-1}$; however, we detected clear radial velocity variations and thus, we conclude that a plausible explanation for the high $v\sin{i}$ is that the star has been spun up by tidal interactions. For KIC~7732964 we derived a $v\sin{i}$ of $23_{-2}^{+1}\,{\rm km\,s}^{-1}$ and a $v_{\rm eq}$ of $30\pm2\,{\rm km\,s}^{-1}$, which is very high considering that this star lies near the base of the RGB. Furthermore, as reported by \citet{howell2016}, it exhibits strong and complex H$\alpha$ and Ca~{\sc ii} H and K emission (Fig. \ref{fig:emission}). Although KIC~7732964 shares similar characteristics with FK~Com (albeit less extreme), it was observed only once in this study and therefore the presence of a binary companion cannot be ruled out.

\begin{table*}
	\caption{The defining properties of known FK~Comae stars (FK~Com, ET~Dra, and HD~199178) compared with those of the 8 stars in our sample. Columns 2 and 3 specify whether the star has evolved off of the MS or if it has a short-period companion. Columns 4 and 5 specify whether X-ray emission or optical emission (Ca~{\sc ii}~K, Ca~{\sc ii}~H, and/or H$\alpha$) has been detected. Columns 6 and 7 list each star's $v\sin{i}$ value and estimated mass.}
	\label{tbl:fk_comae}
	\begin{center}
	\begin{tabular}{@{\extracolsep{\fill}}l c c c c c r@{\extracolsep{\fill}}}
		\hline
		\hline
		\noalign{\vskip0.5mm}
ID & Post-MS & Short-$P_{\rm orb}$ & X-ray & Optical & $v\sin{i}$           &         $M$ \\
   &         & Comp.               & Em.   & Em.     & $({\rm km\,s^{-1}})$ & $(M_\odot)$ \\
		\noalign{\vskip0.5mm}
		\hline
		\noalign{\vskip0.5mm}
FK~Com         & Yes &  No & Yes & Yes &   160$^{\rm a\,}$ &        2.2$^{\rm b\,}$ \\
ET~Dra         & Yes &  No & Yes & Yes & 30-40$^{\rm c\,}$ & $\approx1.4^{\dagger}$ \\
HD~199178      & Yes &  No & Yes & Yes &    80$^{\rm d\,}$ &        2.2$^{\rm b\,}$ \\
\hline
EPIC~210355746 &  No &  No & Yes & Yes &          21 & 0.97 \\
EPIC~210384590 &  No &   ? & Yes & Yes &           9 & 0.86 \\
EPIC~210426551 &  No &   ? & Yes & Yes &         209 & 1.05 \\
EPIC~210505125 &  No & Yes & Yes &  No &   11.7/12.8 & 1.04 \\
EPIC~220533366 & Yes & Yes & Yes & Yes &          17 &  1.2 \\
KIC~4857678    &  No &  No & Yes &  No &         111 & 1.29 \\
KIC~6150124    & Yes &   ? & Yes &  No &           4 &  2.2 \\
KIC~7732964    & Yes &   ? & Yes & Yes &          23 & 0.77 \\
		\noalign{\vskip0.5mm}
		\hline \\
\noalign{\vskip-0.3cm}
\multicolumn{7}{l}{$^{\rm a\,}$\citet{huenemoerder1993}, $^{\rm b\,}$\citet{gondoin1999}, $^{\rm c\,}$\citet{silva1985},}\\
\multicolumn{7}{l}{$^{\rm d\,}$\citet{huenemoerder1986}, $^\dagger$Estimated using \textit{Gaia}~DR2 $T_{\rm eff}$ and $L$ values and }\\
\multicolumn{7}{l}{solar-metallicity \citet{ekstrom2012} evolutionary models.} \\
	\end{tabular}
	\end{center}
\end{table*}

The RGB star KIC~6150124 exhibits the lowest $v\sin{i}$ value ($4\pm1\,{\rm km\,s}^{-1}$) within our sample. Along with the solar-type oscillations that are present in the \emph{Kepler} light curve \citep[$f\sim3.5\,{\rm c\,d}^{-1}$,][]{yu2018}, we identified a possible rotational modulation signature corresponding to $P_{\rm rot}=31.3698\pm0.0001\,{\rm d}$. The estimated $P_{\rm rot}$ and $R$ values imply an unusually high equatorial velocity of $19.7\pm0.8\,{\rm km\,s}^{-1}$; however, due to the increased low-frequency power produced by granulation, the estimated $P_{\rm rot}$ is not considered to be reliable. We conclude that this star is likely not a rapid rotator.

Two of the 8 targets in our study may be classified as ultra rapid rotators based on their extremely high $v\sin{i}$ values. For KIC~4857678, we derive a $v\sin{i}$ value of $111_{-1}^{+6}\,{\rm km\,s}^{-1}$ ($\approx34$~per~cent of $v_{\rm crit}$) while no rotation period could be estimated from its \emph{Kepler} light curve. For EPIC~210426551, we derive a $v\sin{i}$ value of $209\pm13\,{\rm km\,s}^{-1}$ ($\approx68$~per~cent of $v_{\rm crit}$) and a photometrically-inferred equatorial velocity of $184\pm12\,{\rm km\,s}^{-1}$, which implies that the rotation axis has an inclination angle $\sim90\,{\rm degrees}$. These two stars are discussed below.

KIC~4857678 has a fractional MS age of $0.7$ estimated using evolutionary models \citep{ekstrom2012,georgy2013}. No Zeeman signatures were detected in the LSD profiles (we obtained a minimum precision of $\sigma_{\langle B_z\rangle}=64\,{\rm G}$) and no H$\alpha$ or Ca~{\sc ii}~H and K emission was detected suggesting that the star is not magnetically active. The transition between those MS stars with deep convective envelopes and those with shallow convective envelopes is predicted to occur near $\approx1.4\,M_\odot$ \citep[e.g.][]{vansaders2012,cantiello2019}; this boundary approximately coincides with the gradual increase in frequency of strongly magnetic A-type (Ap) stars \citep[e.g. Fig. 17 of][]{sikora2019}, which account for $\lesssim10$~per~cent of all MS A-type stars. Based on KIC~4857678's derived mass of $1.29^{+0.05}_{-0.07}\,M_\odot$, it is plausible that its convective envelope is shallow and unable to generate magnetic fields that might otherwise enable significant spin-down via magnetic braking to take place. The fact that we do not detect any spectroscopic evidence of magnetic activity appears to conflict with the strong X-ray and UV emission that was detected for this star \citep{smith2015}. No evidence of a short-period binary companion was found based on the 2 observations that were obtained. Due to the relatively low angular resolution of the X-ray and UV instruments, the detected emission may be associated with a widely separated magnetically active companion (e.g. a late M-type star) or a hot white dwarf \citep[e.g.][]{heise1985}. Ultimately, we conclude that this star's apparent rapid rotation rate is likely consistent with the current predictions of single star evolution.

As with KIC~4857678, we did not detect Zeeman signatures for EPIC~210426551 although this may be due to the large uncertainties ($\sigma_{\langle B_z\rangle}=335\,{\rm G}$) that were obtained primarily as a result of the star's high $v\sin{i}$ value. Comparisons between the observed spectra and the model spectra generated using the best-fitting parameters do yield possible evidence of magnetic activity via weak Ca~{\sc ii}~H and K emission and possibly weak H$\alpha$ emission (Fig. \ref{fig:emission}). Several examples of young, magnetically active solar-type stars with $v\sin{i}$ values $>200\,{\rm km\,s}^{-1}$ have been previously reported \citep[e.g.][]{marsden2009a}. Under the assumption that magnetic activity ought to increase with increasing rotational velocity, one might expect spin-down to occur more rapidly in these cases, thereby preventing such rapid rotators from appearing near the end of the pre-MS phase. It has been proposed that these observations may be explained by the fact that the efficiency of magnetic braking does not continue to increase with increasing rotational velocity above a certain saturation limit \citep[e.g.][]{vilhu1984,krishnamurthi1997,wright2011}. Models of the rotational evolution of solar-type stars that incorporate a saturation threshold have successfully replicated the observed convergence towards low rotational velocities ($\lesssim10\,{\rm km\,s}^{-1}$) that occurs after $\sim1\,{\rm Gyr}$ \citep[e.g.][]{spada2011,amard2016}.

Based on the preceding discussion, it may only be possible to explain EPIC~210426551's extremely high rotational velocity as the product of a single star evolutionary history if it is very young (i.e. either a ZAMS or pre-MS star). For EPIC~210426551, we derive a metallicity of [M/H]$=-0.4\pm0.1$, which permits two possibilities in terms of its evolutionary state: either it is close to the terminal-age MS (TAMS) (see Fig. \ref{fig:HRD}) or it is a pre-MS star approaching the ZAMS. Comparisons between the \emph{Gaia} DR2 astrometry, proper motion, and distance measurements with those of nearby open clusters \citep{kharchenko2013} did not yield any clear potential memberships that could have otherwise been used to infer age constraints. One possible indication of youth is the detection of excess IR flux, which is generally indicative of dust particles \citep[e.g.][]{wyatt2008}. Using the criterion proposed by \citet{wu2013} based on 2MASS and WISE photometry \citep{cohen2003,wright2010}, we find no indication of an IR excess for EPIC~210426551.

We note that if the derived metallicity for EPIC~210426551 is inaccurate and it instead has a solar metallicity, then it is expected to have a similar age and mass to KIC~4857678. In this scenario, the star's high rotation rate can be explained either by the fact that (1) the star is not subject to significant magnetic braking due to a shallow or non-existant convective envelope (similar to the scenario proposed for KIC~4857678) or (2) that the star is very young (either ZAMS or pre-MS). Considering that we have detected evidence of magnetic activity (as noted above; see Fig. \ref{fig:emission}), it appears most likely that if EPIC~210426551 has a solar metallicity, the latter of these two explanations is more plausible. Of course, only a single observation was obtained in this study and we cannot yet rule out the possibility that this star has been spun-up by binary interactions (N.B. the `duplicated\_source' flag in the \emph{Gaia} DR2 catalogue is `false' implying that no additional sources associated with EPIC~210426551 were detected). Therefore, without having better constraints on either this object's age or whether it has a short-period binary companion, we conclude that the explanation for EPIC~210426551's high rotational velocity as being the product of a stellar merger event remains viable.

Stellar merger events are predicted to occur largely as a result of primordial binary evolution \citep[e.g.][]{webbink1976a,politano2010,howitt2020}. Theoretical predictions related to this formation channel have suggested that the merging of the two stars may be facilitated by the presence of a widely separated third component via Kozai-Lidov cycles \citep{kozai1962,lidov1962,portegieszwart2016,izzard2018}. It is therefore interesting to note that, in the case of the 2 targets identified here as potential stellar merger candidates (KIC~7732964 and EPIC~210426551), high-resolution speckle interferometric measurements have yielded detections of wide binary companions ($P_{\rm orb}\gtrsim17\,{\rm yrs}$) (Howell et al., in prep.). Such detections, if shown to be related to stellar mergers, may provide a further identifying criterion in addition to those previously discussed (e.g. rapidly rotating, magnetically active evolved stars).

\section{Summary}\label{sect:summ}

We have obtained 1-2 high-resolution spectropolarimetric (Stokes $V$) observations for a small subset (8 targets) of a larger sample of candidate FK~Comae type stars that were first identified by \citet{smith2015} and \citet{howell2016}. We conclude that none of the 8 targets are FK~Comae candidates based on (1) the detection of binary companions, (2) their inferred evolutionary states, or (3) low rotational velocities ($<10\,{\rm km\,s}^{-1}$) inferred from spectroscopically-derived $v\sin{i}$ values and/or photometrically-inferred rotation periods. This is not particularly surprising considering the extreme rarity of FK~Comae type stars. However, 2 targets included in our study (KIC~7732964 and EPIC~210426551) are found to exhibit characteristics that are consistent with some of the predicted properties of stellar merger products (i.e. the leading explanation for FK~Com's unusual characteristics) and, based on the currently available observations, they cannot be ruled out as such. At a minimum, additional spectra are required for these 2 targets in order to search for binary companions that may explain their anomalously high rotational velocities. If no short-period binary companions are found to be associated with these stars, obtaining better constraints on (1) their apparent magnetic activity using instruments such as ESPaDOnS@CFHT or SPIRou@CFHT or on (2) the presence of cool dust using instruments provided for example by SOFIA, may yield important clues about the nature of these objects and stellar merger products in general.

\section*{Acknowledgments}

GAW acknowledges support in the form of a Discovery Grant from the Natural Science and Engineering Research Council (NSERC) of Canada.

\bibliography{stellar_mergers}
\bibliographystyle{mn2e}

\clearpage

\section*{Appendix}

\begin{figure}
	\centering
	\subfigure{\includegraphics[width=1\columnwidth]{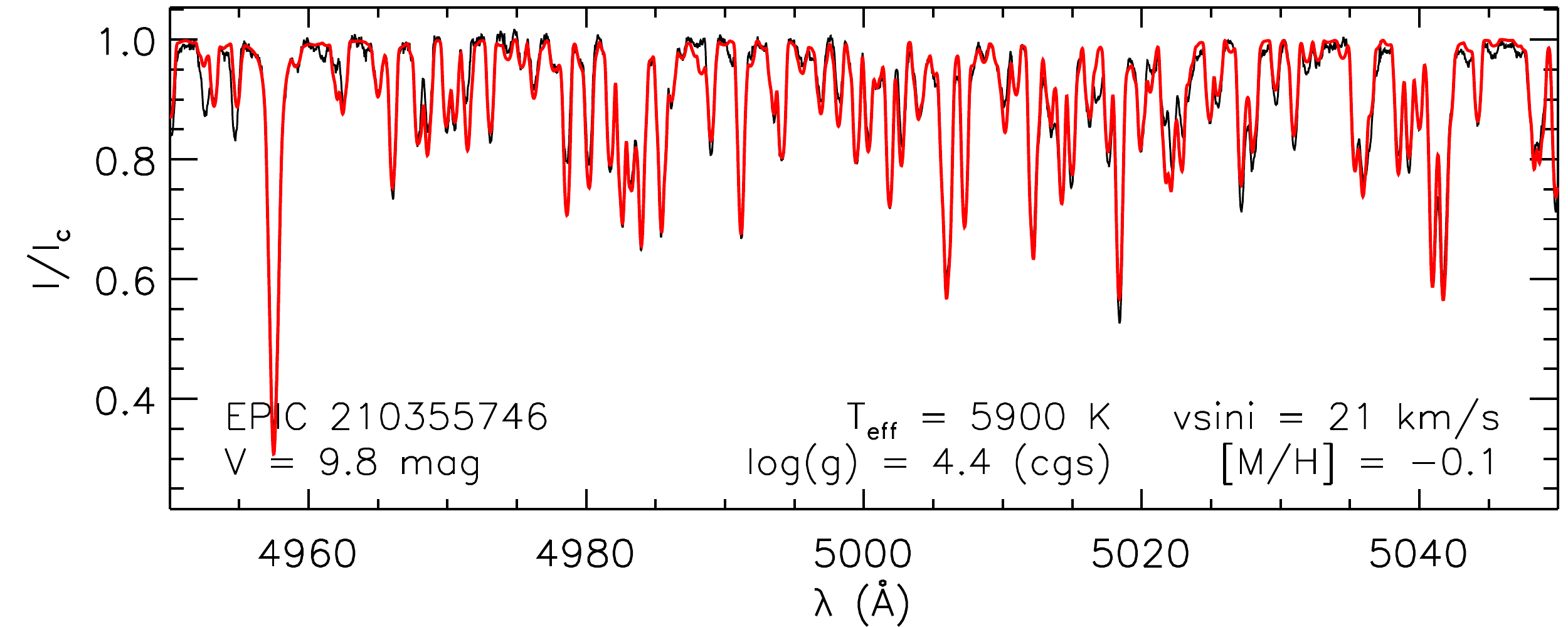}}\vspace{-0.4cm}
	\subfigure{\includegraphics[width=1\columnwidth]{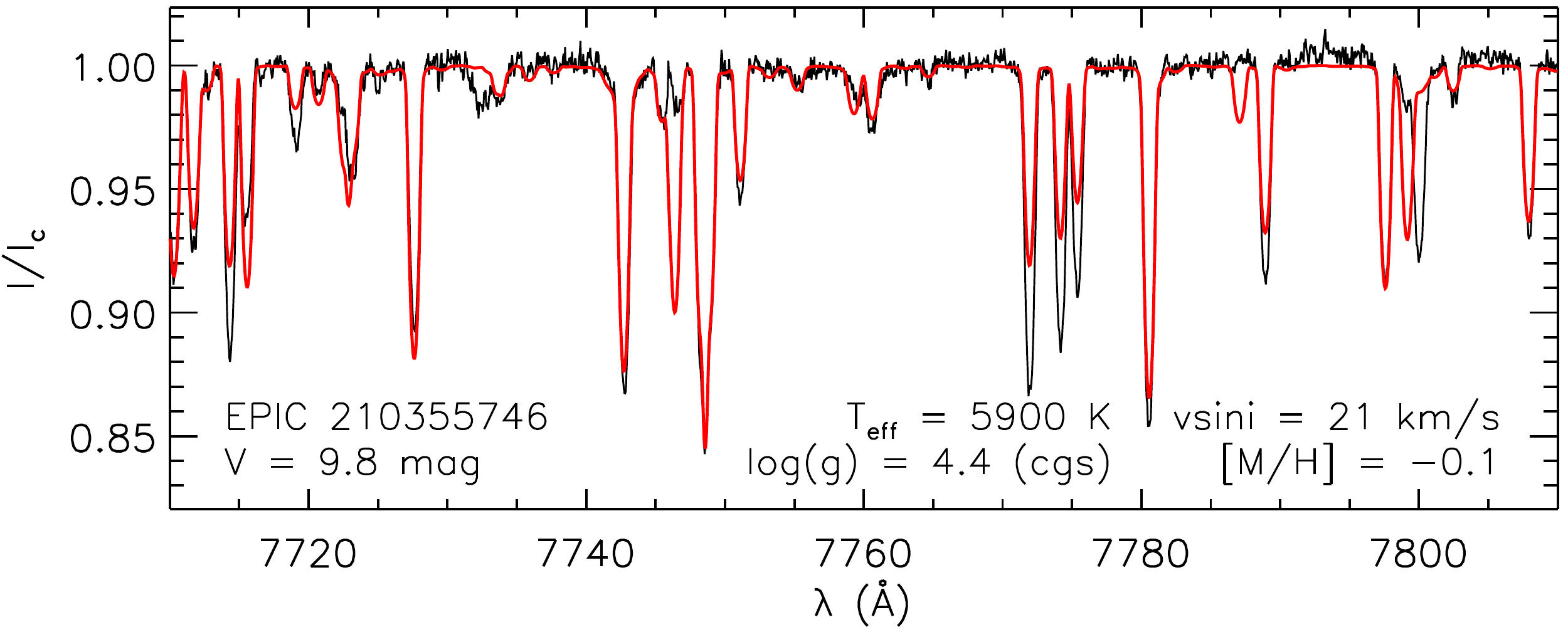}}\vspace{-0.4cm}
	\subfigure{\includegraphics[width=1\columnwidth]{EPIC210355746_avg_8700_8825-eps-converted-to.pdf}}\vspace{-0.4cm}
	\caption{Comparisons between the model spectra generated using the best-fitting parameters (red curve) and the observed spectra (black curve) for the 3 selected spectral windows ($4\,950-5\,050\,{\rm \AA}$, $7\,710-7\,810\,{\rm \AA}$, and $8\,700-8\,825\,{\rm \AA}$).}
	\label{fig:spec_fit01}
\end{figure}

\begin{figure}
	\centering
	\subfigure{\includegraphics[width=1\columnwidth]{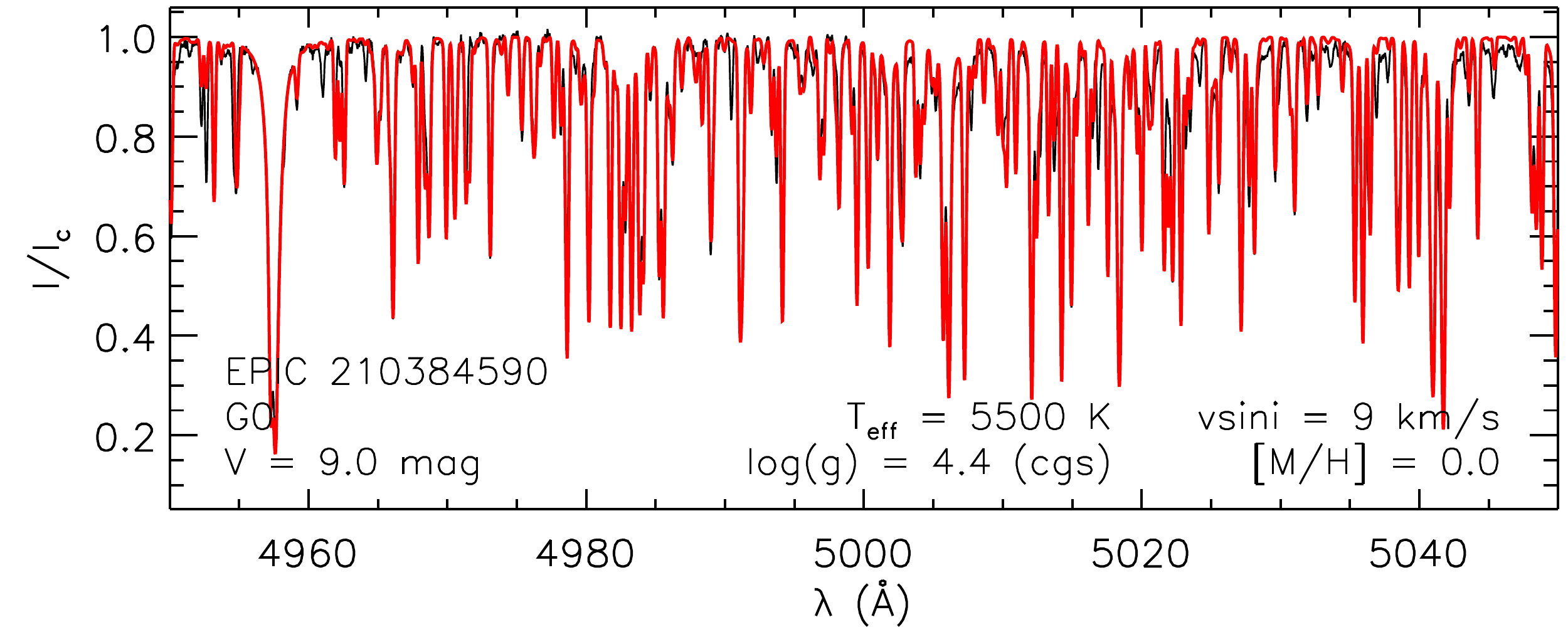}}\vspace{-0.4cm}
	\subfigure{\includegraphics[width=1\columnwidth]{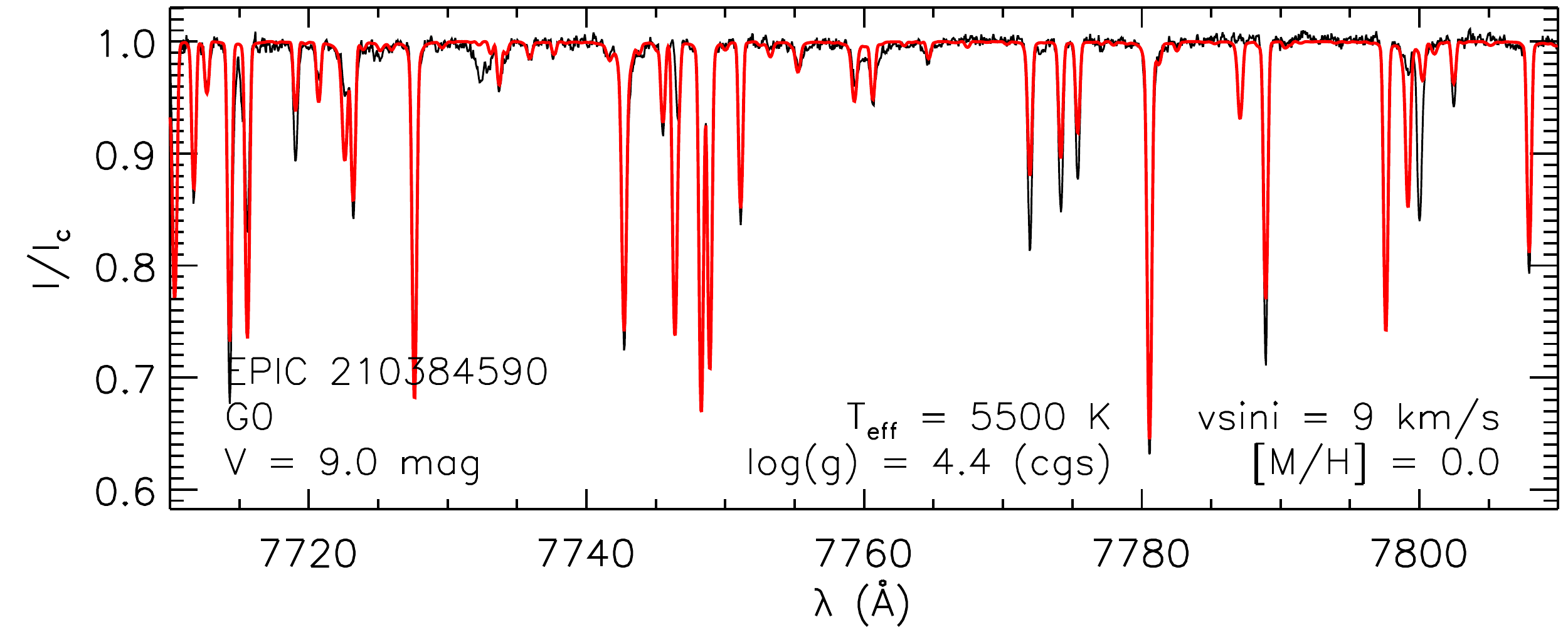}}\vspace{-0.4cm}
	\subfigure{\includegraphics[width=1\columnwidth]{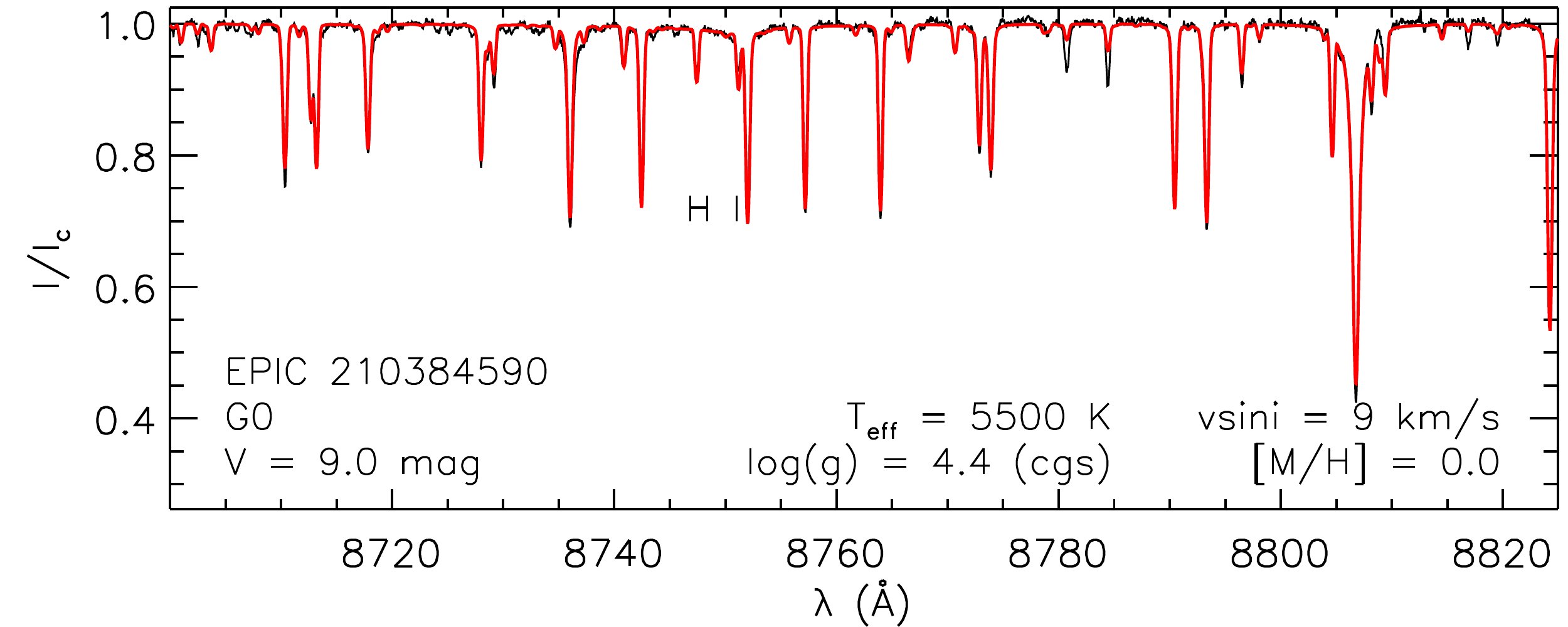}}\vspace{-0.4cm}
	\caption{Same as Fig. \ref{fig:spec_fit01}.}
	\label{fig:spec_fit02}
\end{figure}

\begin{figure}
	\centering
	\subfigure{\includegraphics[width=1\columnwidth]{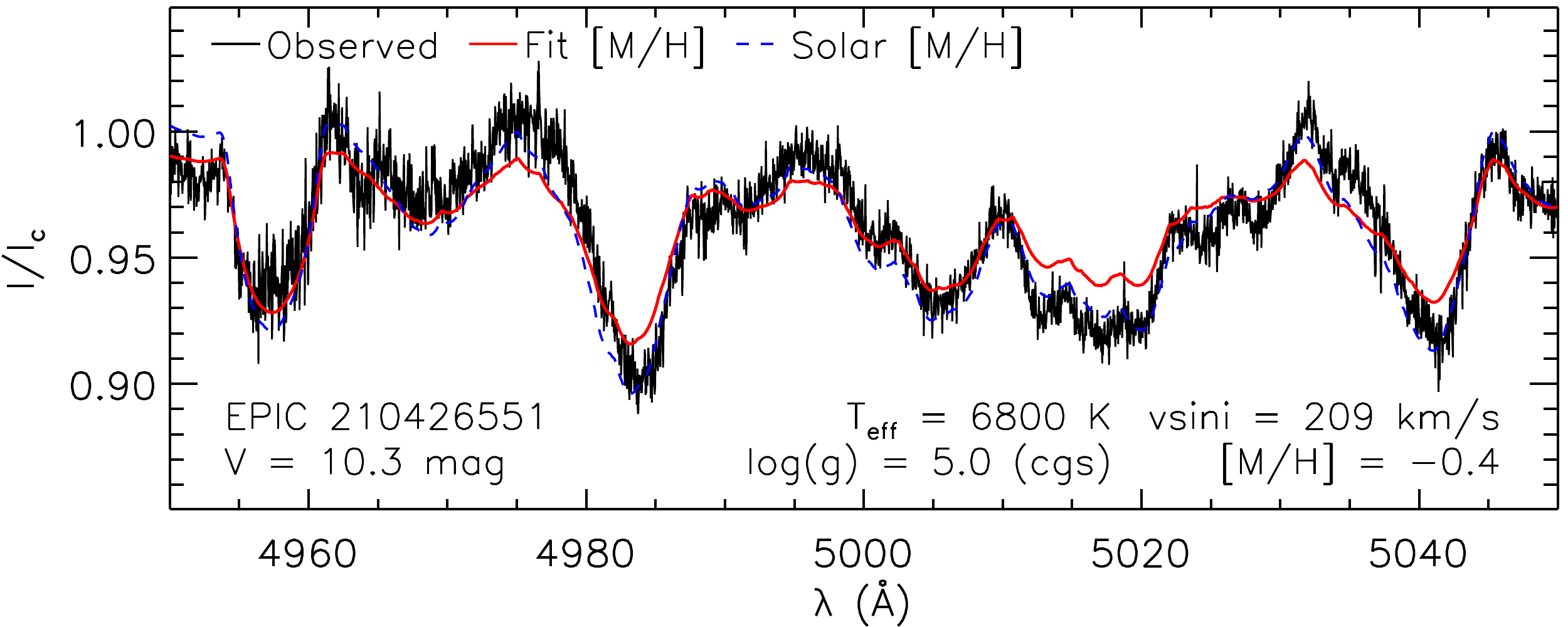}}\vspace{-0.4cm}
	\subfigure{\includegraphics[width=1\columnwidth]{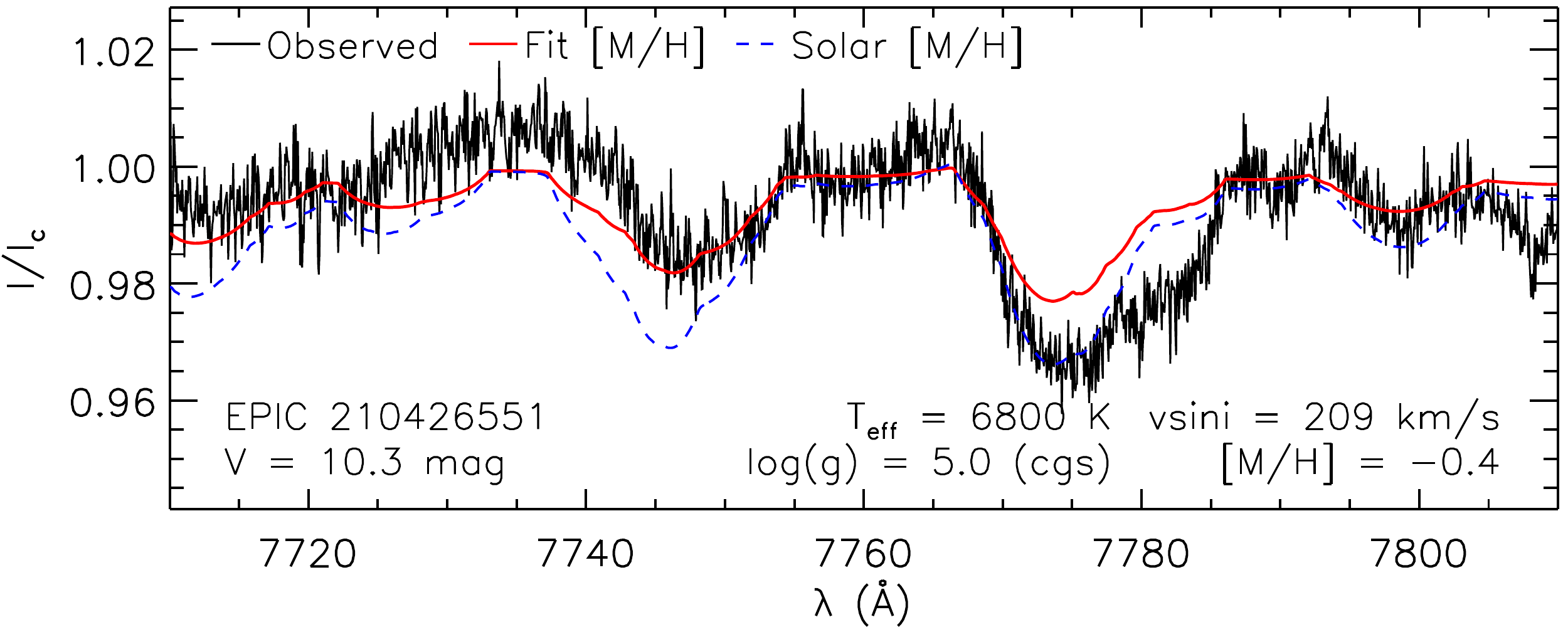}}\vspace{-0.4cm}
	\subfigure{\includegraphics[width=1\columnwidth]{EPIC210426551_avg_8700_8825-eps-converted-to.pdf}}\vspace{-0.4cm}
	\caption{Same as Fig. \ref{fig:spec_fit01}. The dashed blue curve corresponds to the model spectra generated using the best-fitting parameters but with a solar metallicity ([M/H]$=0$).}
	\label{fig:spec_fit03}
\end{figure}

\begin{figure}
	\centering
	\subfigure{\includegraphics[width=1\columnwidth]{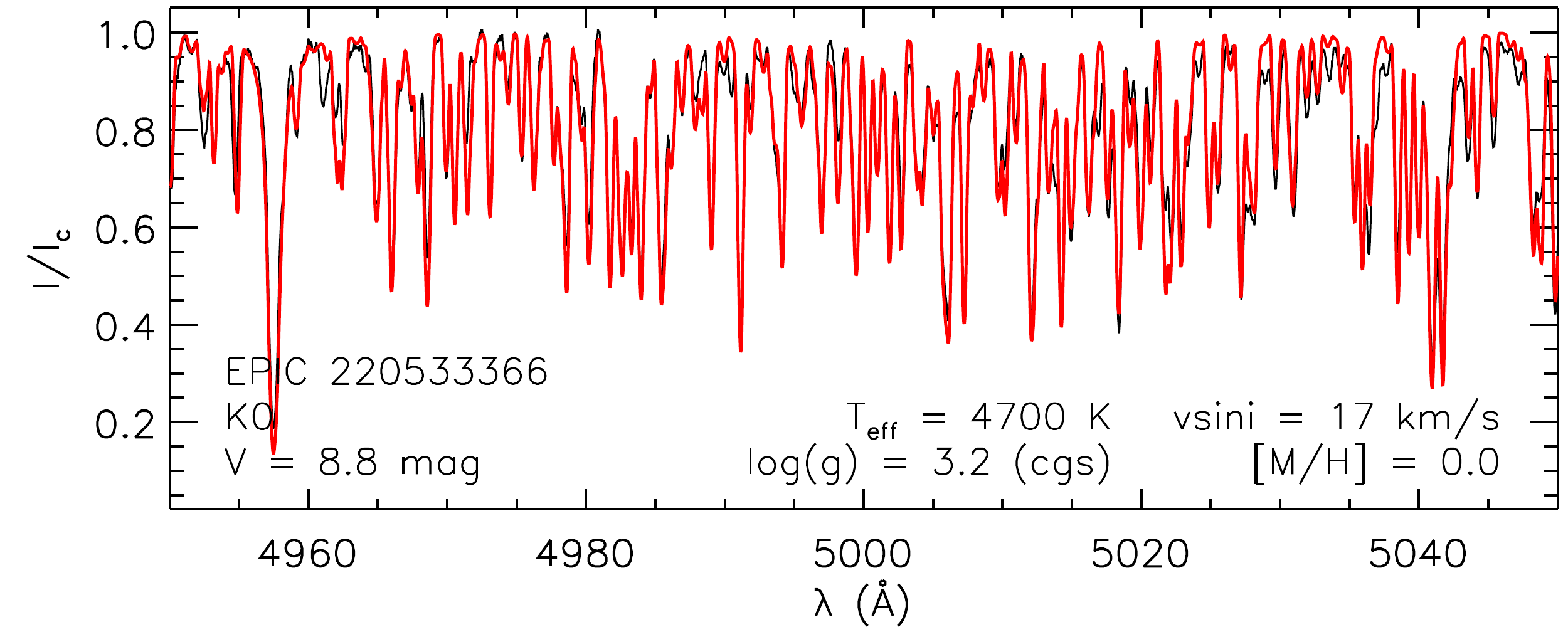}}\vspace{-0.4cm}
	\subfigure{\includegraphics[width=1\columnwidth]{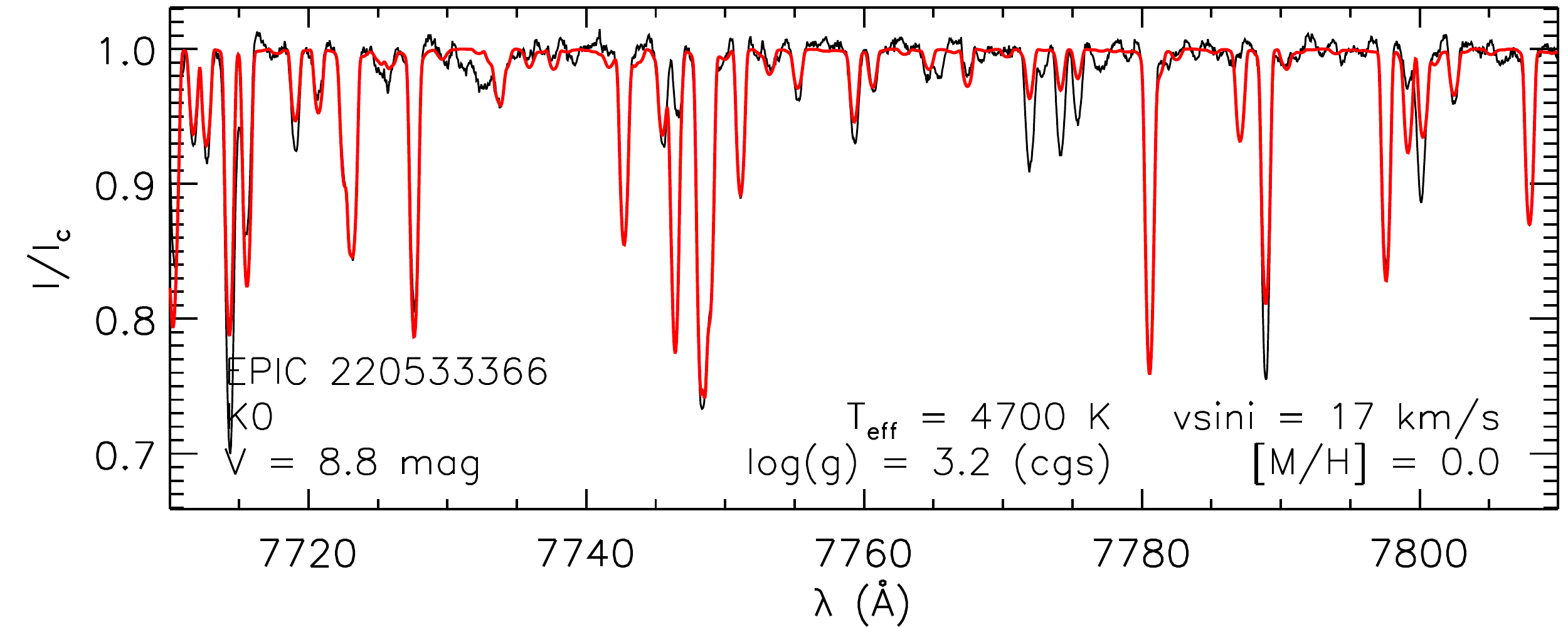}}\vspace{-0.4cm}
	\subfigure{\includegraphics[width=1\columnwidth]{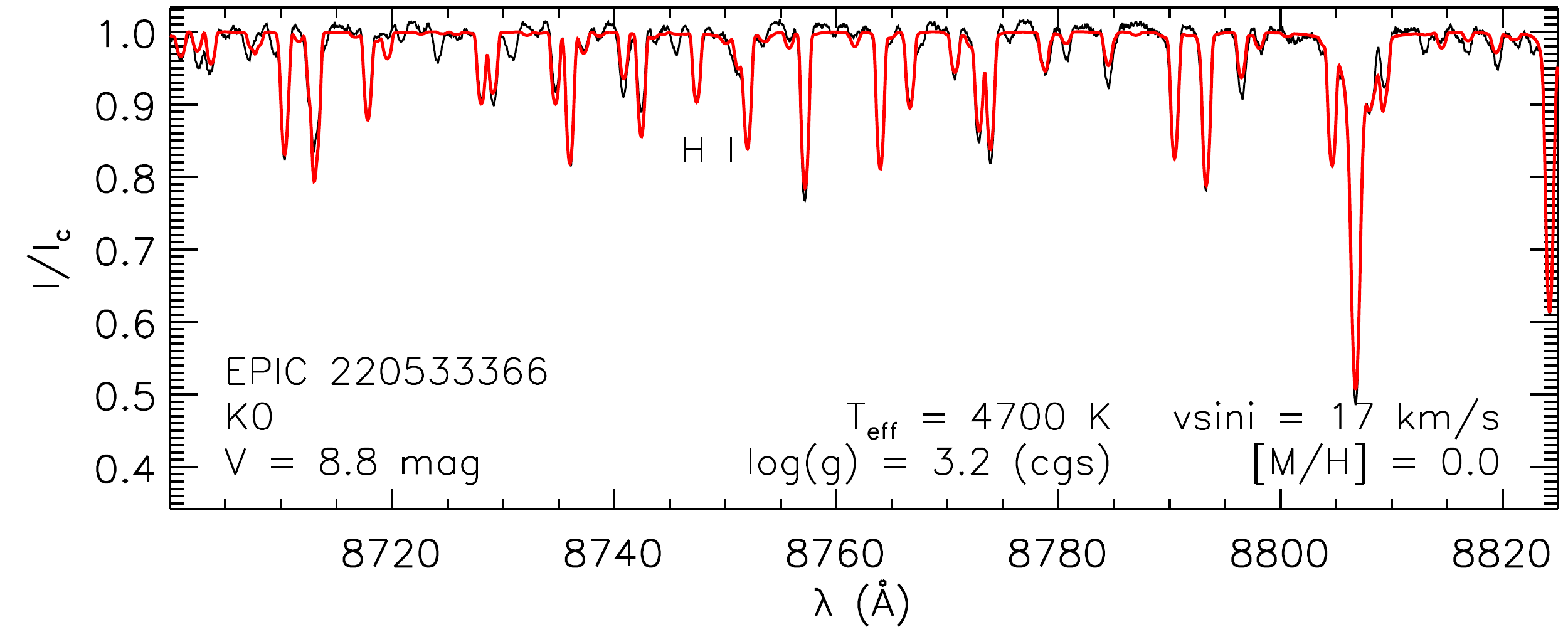}}\vspace{-0.4cm}
	\caption{Same as Fig. \ref{fig:spec_fit01}.}
	\label{fig:spec_fit04}
\end{figure}

\begin{figure}
	\centering
	\subfigure{\includegraphics[width=1\columnwidth]{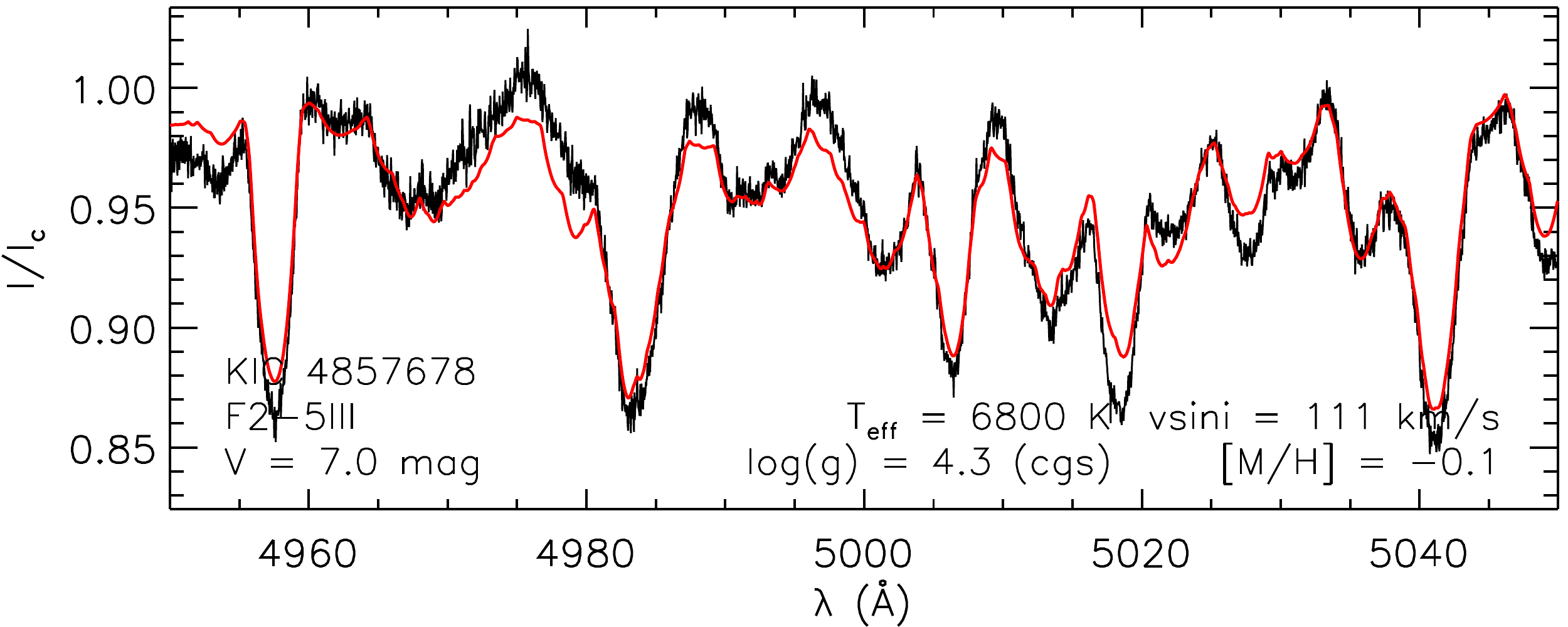}}\vspace{-0.4cm}
	\subfigure{\includegraphics[width=1\columnwidth]{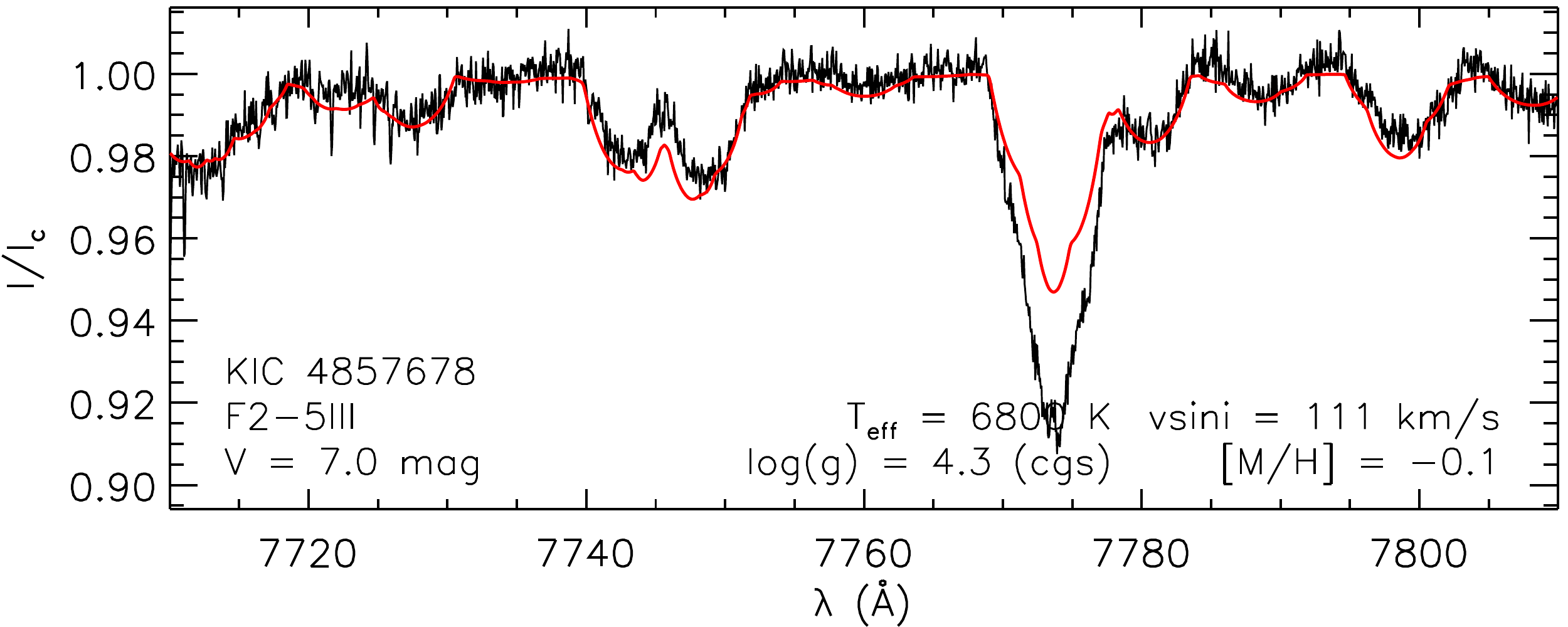}}\vspace{-0.4cm}
	\subfigure{\includegraphics[width=1\columnwidth]{KIC04857678_avg_8700_8825-eps-converted-to.pdf}}\vspace{-0.4cm}
	\caption{Same as Fig. \ref{fig:spec_fit01}.}
	\label{fig:spec_fit05}
\end{figure}

\begin{figure}
	\centering
	\subfigure{\includegraphics[width=1\columnwidth]{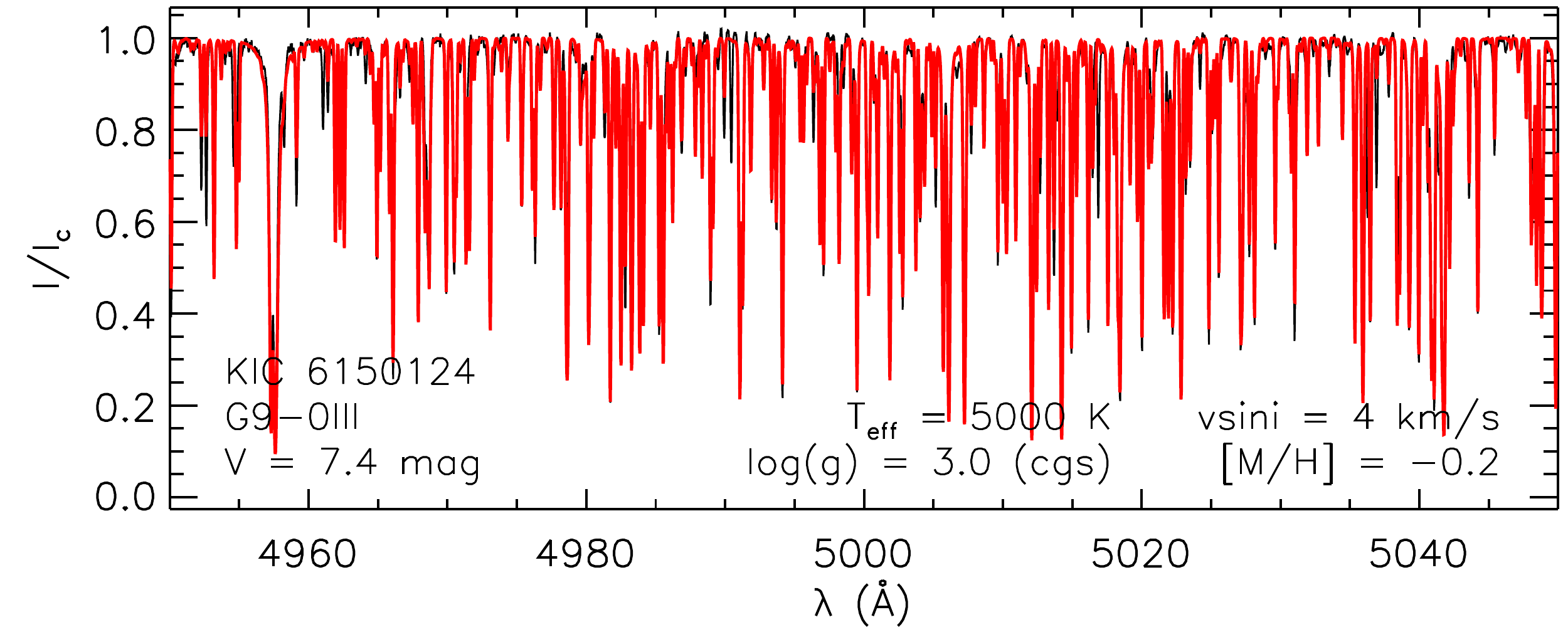}}\vspace{-0.4cm}
	\subfigure{\includegraphics[width=1\columnwidth]{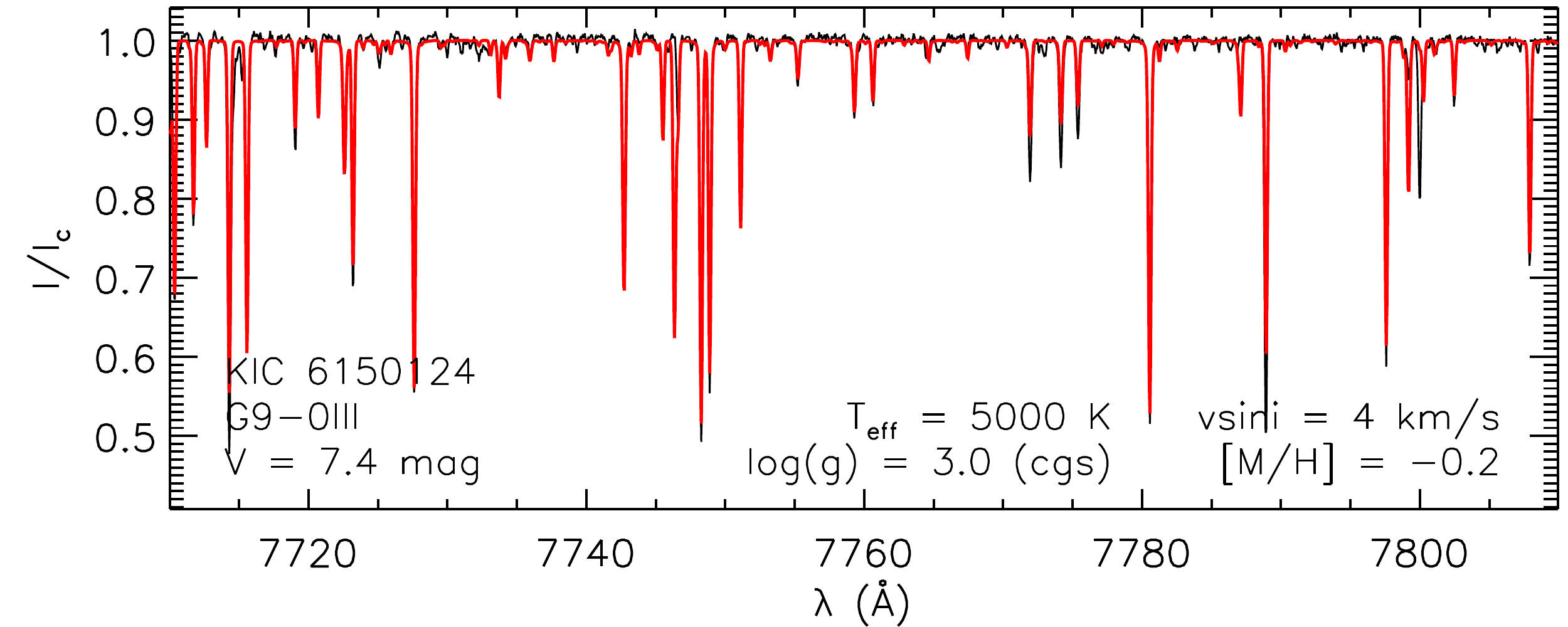}}\vspace{-0.4cm}
	\subfigure{\includegraphics[width=1\columnwidth]{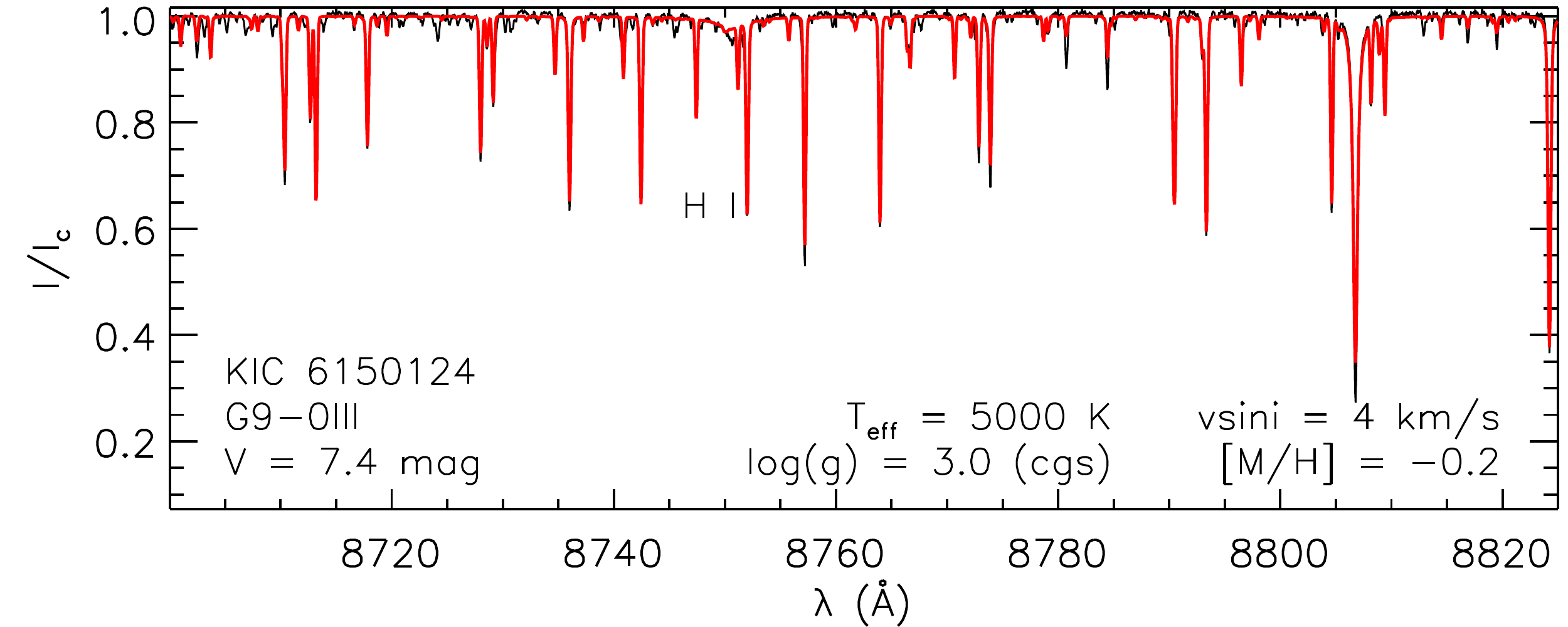}}\vspace{-0.4cm}
	\caption{Same as Fig. \ref{fig:spec_fit01}.}
	\label{fig:spec_fit06}
\end{figure}

\begin{figure}
	\centering
	\subfigure{\includegraphics[width=1\columnwidth]{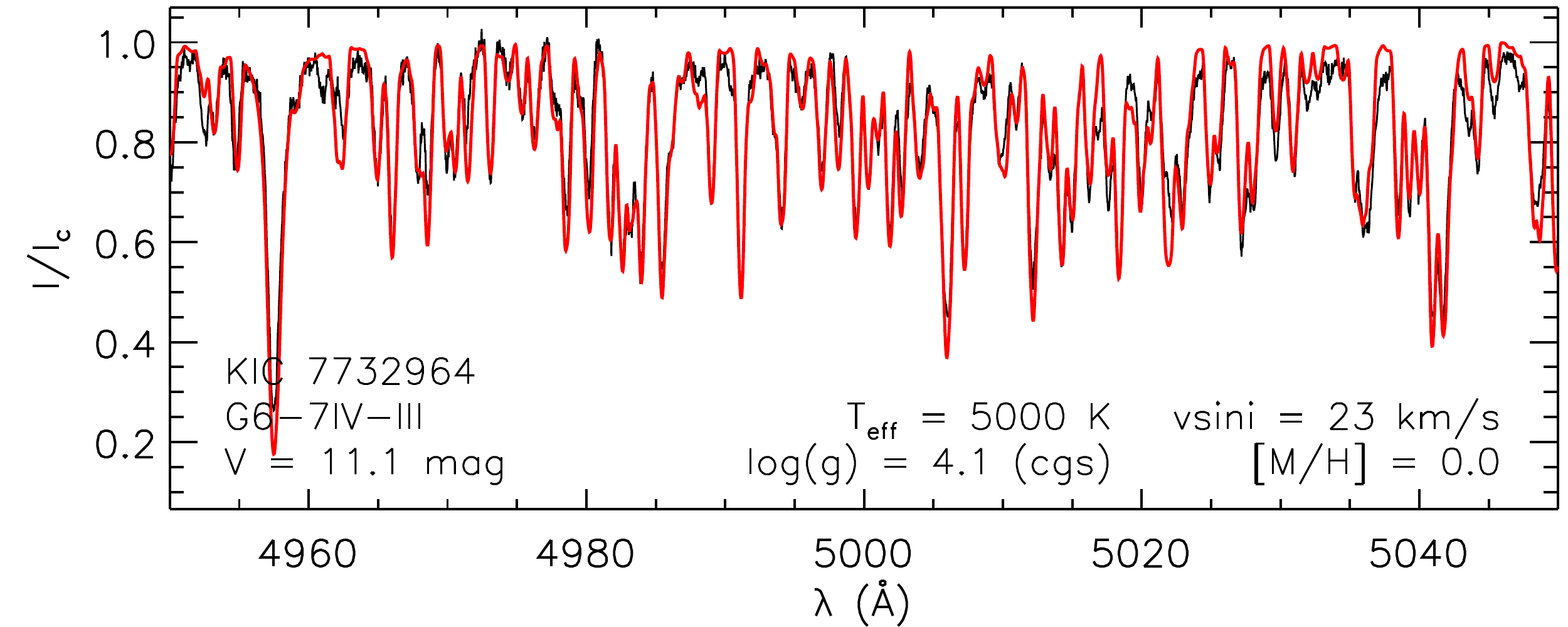}}\vspace{-0.4cm}
	\subfigure{\includegraphics[width=1\columnwidth]{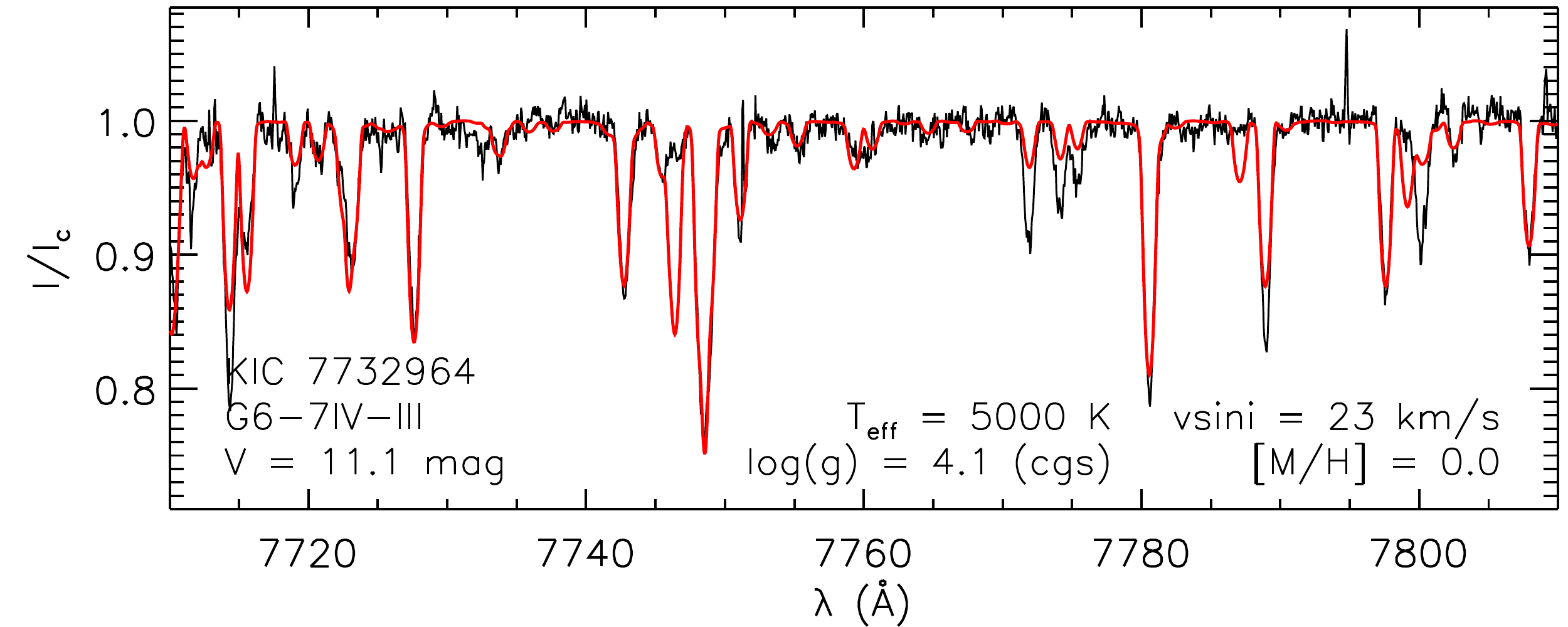}}\vspace{-0.4cm}
	\subfigure{\includegraphics[width=1\columnwidth]{KIC07732964_avg_8700_8825-eps-converted-to.pdf}}\vspace{-0.4cm}
	\caption{Same as Fig. \ref{fig:spec_fit01}.}
	\label{fig:spec_fit07}
\end{figure}

\end{document}